\documentclass[12pt]{article}
\usepackage{amsthm}
\usepackage{amsmath}
\usepackage[colorlinks,citecolor=blue,urlcolor=blue,filecolor=blue,backref=page]{hyperref}
\usepackage[english]{babel}
\usepackage[utf8]{inputenc}
\usepackage{amsmath}
\usepackage{amssymb}
\usepackage{comment}
\usepackage{mathrsfs}
\usepackage{amsthm}
\usepackage{bbm}
\usepackage{rotating}
\usepackage{lscape}
\usepackage{subfigure}
\usepackage[colorinlistoftodos]{todonotes}
\usepackage{lipsum}
\usepackage{comment}
\usepackage{graphicx}
\usepackage{natbib}
\usepackage{url}
\usepackage{graphicx}

\numberwithin{equation}{section}

\newcommand{\ignore}[1]{}

\newcommand{\sF}{\mathcal{F}}
\newcommand{\sG}{\mathcal{G}}

\newcommand{\IE}{\mathbb{E}}

\newcommand{\blind}{0}

\begin{document}
	
	\def\spacingset#1{\renewcommand{\baselinestretch}%
		{#1}\small\normalsize} \spacingset{1}

	\if0\blind
	{
		\title{\bf {Analyzing initial stage of COVID-19 transmission through Bayesian time-varying model}}
		\author{Arkaprava Roy, 
			Sayar Karmakar\\
			University of Florida}
		\maketitle
	} \fi
	
	\if1\blind
	{
		\bigskip
		\bigskip
		\bigskip
		\begin{center}
			{\LARGE\bf }
		\end{center}
		\medskip
	} \fi
	
	\bigskip
	
	\begin{abstract}
		Recent outbreak of the novel coronavirus COVID-19 has affected all of our lives in one way or the other. While medical researchers are working hard to find a cure and doctors/nurses to attend the affected individuals, measures such as `lockdown', `stay-at-home', `social distancing' are being implemented in different parts of the world to curb its further spread. To model the non-stationary spread, we propose a novel time-varying semiparametric AR$(p)$ model for the count valued time-series of newly affected cases, collected every day and also extend it to propose a novel time-varying INGARCH model. Our proposed structures of the models are amenable to Hamiltonian Monte Carlo (HMC) sampling for efficient computation. We substantiate our methods by simulations that show superiority compared to some of the close existing methods. Finally we analyze the daily time series data of newly confirmed cases to study its spread through different government interventions.
	\end{abstract}
	
	\noindent%
	{\it Keywords:}  Autoregressive model, B-splines, COVID-19, Count-valued time series, Hamiltonian Monte Carlo (HMC), INGARCH, Non-stationary, Poisson Regression 
	
	\spacingset{1.45}
	
	
	\section{Introduction}
	
	Coronavirus is a class of viruses that primarily affect mammals and birds. The viruses in this class predominantly cause respiratory infections among humans. Most of these viruses from this class only cause mild respiratory infections or the common cold. To date, three viruses in this class have been turned out to be deadly. In 2002-03, there was an outbreak of Severe Acute Respiratory Syndrome (SARS) with 11\% fatality rate \citep{chan2003sars}. The year 2015 observed another deadly coronavirus Middle East Respiratory Syndrome(MERS) with 35\% fatality\citep{alsolamy2015infection}. The third one is COVID-19 which has caused this current outbreak. The reported fatality of this virus is yet very low as compared to the other two. However, it spreads much faster and can cause ``community spread" when the cause of the infection can no longer be traced back to its source. The source of this virus has been traced back to the wet market in Wuhan, China and dates back to December 2019. Since then, it has been spreading across the world. On January 20, 2020, United States (USA) recorded its first COVID-19 patient in a man, returning from Wuhan, China. Italy reported its first confirmed case on January 31, 2020. After that, it has been spreading continuously.
	
	Since there is no vaccine yet in the market, the government is implementing strong measures such as city-wide or state-wide `lockdown's, `stay-at-home' notices, extensive testing to control the outbreak. Thus, this count-valued time-series of daily new cases of infection is expected to vary largely. While initially by the contagious nature these spread looks exponential, once some measures are undertaken the number of cases goes down but at varying degrees depending on how strict was the enforcement, how dense is the population, etc. There has been ample research study in a very short time to discuss the effectiveness of lockdown and forecast of the future path of how the virus will spread. But somehow a comprehensive understanding of the statistical model, its estimation and uncertainty quantification remains inadequate. SIR model \citep{song2020epidemiological}, proportional model \citep{deb2020}, some Bayesian epidemic models \citep{clancy2008bayesian, jewell2009bayesian} etc. have been used in the context of continuous modelling because it benefits time-series formulation and ready computation. Relatively straight-forward models have been considered such as polynomial trends or presence of an ARMA structure etc. However, we wish to stick to the actual daily count of new affections and this brings us to a unique juncture of analyzing a count time series with smooth varying coefficients.
	
	Modeling count time series is important in many different fields such as disease incidence, accident rates, integer financial datasets such as price movement, etc. This relatively new research stream was introduced in \cite{zeger1988regression} and interestingly he analyzed another outbreak namely the US 1970 Polio incidence rate. This stream was furthered by \cite{chan1995} where Poisson generalized linear models (GLM) with an autoregressive latent process in the mean are discussed. A wide range of dependence was explored in \cite{davis2003} for simple autoregressive (AR) structure and external covariates. On the other hand, a different stream explored integer-valued time series counts such as ARMA structures as in \citep{brandt2001linear, biswas2009discrete} or INGARCH structure as done in \cite{zhu2011negative, zhu2012zero, zhu2012modeling1,zhu2012modeling}. However, from a Bayesian perspective, only work to best of our knowledge is that of \cite{silveira2015bayesian} where the authors discussed an ARMA model for different count series parameters. However, their treatment of ignoring zero-valued data or putting the MA structure by demeaned Poisson random variable remains questionable. None of these works focused on the time-varying nature of the coefficients except for a brief mention in \cite{sayar2020}.
	
	The rapid change in the observed counts make all earlier time-constant analysis inappropriate and builds a path where we can explore methodological and inferential development in tracking down the trajectory of this spread. Thus, we propose a novel semiparametric time-varying autoregressive model for counts to study the spread and examine the effects of these interventions in the spread based on the time-varying coefficient functions. A time-varying AR$(p)$ process consists of a time-varying mean/intercept function along with time-varying autoregressive coefficient functions. We further generalize it to a time-varying integer-valued generalized autoregressive conditional heteroscedasticity (tvINGARCH) model where the conditional mean depends also on the past conditional means. Given the exponential trend of the spread, it is expected that the mean would vary with the level.
	
	Our goals are motivated by both the application and methodological development. To the best of our knowledge, ours is the first attempt to model possibly autoregressive count time series with time-varying coefficients. The mean function stands for the overall spread and the autoregressive coefficients stand for different lags. Since this virus can be a largely asymptomatic carrier for the first few days we wish to identify which lags are significant in our model which can be directly linked to how many days the symptom spread but did not show up. We show that for different areas lags 6 to 10 are significant. These findings are in-line with several research articles discussing the incubation length for the novel coronavirus with a median of 5-6 days and 98\% below 11 days. For example, see \cite{incubation}. A few provinces, state, countries have ordered lockdown or stay-at-home orders of various degree and we find that even after these orders are in effect it takes about 12-16 days to reach the peak and then the intercept coefficient function starts decreasing. This is also an interesting find which characterizes the fact that the number of infected but asymptomatic cases is large compared to the new cases reported. Additional to the time-varying AR model proposal, we also offer an analysis via time-varying INGARCH model that assumes an additional recursive term in the conditional expectation (cf. \eqref{TVBING}). This extension offers some more comprehensiveness in the modeling part as even INGARCH with small orders can help us get rid of choosing an appropriate maximum lag value. Since for a Poisson model, the mean is same as the variance, this can also be thought as an extension of the GARCH model in the context of count data. First introduced by \cite{ferland2006integer}, these models were thoroughly analyzed in \cite{zhu2012zero,zhu2012modeling1,zhu2011negative, zhu2012modeling, ahmad2016poisson}. Our proposal of time-varying INGARCH model adapts to the non-stationarity theme and also can be viewed as a new contribution. Finally, we contrast the time-varying AR and the GARCH for both simulations and real-data applications under different metrics of evaluation.
	
	An important criticism of the estimate of basic reproduction number in the various research items over the past two months of Feb and March 2020 is its huge degree of variability. This skepticism is natural as the serial interval distribution of a disease cannot be estimated consistently unless we have an exact infector/infectee pair dataset. The majority of this research is pulling out these estimates based on what we know about two other outbreaks, namely SARS and MERS. One can easily see the R0 estimates from the popular R0 package by \cite{R0packagepaper} depends on the start and end date and a plugin value of the serial distribution. Keeping this in mind, we decided to NOT have R0 coefficient in our model. Instead, we focus on the autoregressive coefficient functions and provide insights which we believe can be used to develop new estimates of the basic reproduction number from the data itself, without having to rely on the serial number distribution from other diseases.

	In our present context, the number of affected can be covered by the popular SIR model in this case, however, they assume additional structure on how these numbers evolve and then tries to estimate the rate. Instead, we do not assume any such specific evolution and offer a general perspective. Our simulation results corroborate a consistent estimation of the unknown functions. Regression models with varying coefficient were introduced by 
	\cite{hastie1993varying}. They modeled the varying coefficients using cubic B-splines. Later, these models has been further explored in various directions \cite{gu1993smoothing,biller2001bayesian,fan2008statistical,franco2019unified,yue2014bayesian}. Spline bases have been routinely used to model the time-varying coefficients within non-linear time series models \citep{cai2000functional, huang2002varying, huang2004functional,amorim2008regression}. We also consider the B-spline series based priors to model the time-varying coefficients in our model. In Bayesian literature of non-linear function modeling, B-splines series based priors have been extensively developed under different shape constraints \citep{he1998monotone,meyer2012constrained, das2017bayesian,mulgrave2018bayesian, roy2018high}. To the best of our knowledge, there is no other work that puts \eqref{eq:parcondition} or \eqref{eq:parconditionG}-specific shape constraints, required for a time-varying AR or GARCH using B-spline series. We also discuss a pointwise inferential tool by drawing credible intervals. Such tools are important to keep an objective perspective in terms of the evolution of the time-varying coefficients without restricting it to some specific trend models. See \cite{sayar2020} ( \cite{karmakar2018asymptotic} for an earlier version) for a comprehensive discussion on time-varying models and their applications. 
	
	The rest of the paper is organized as follows. Section \ref{model} describes the proposed Bayesian models in detail. Section~\ref{comp} discusses an efficient computational scheme for the proposed method. We study the performance of our proposed method in capturing true coefficient functions and show excellent performance over other existing methods in Section~\ref{sim}. Section~\ref{application} deals with an application of the proposed method on COVID-19 spread for different countries. Then, we end with discussions, concluding remarks and possible future directions in Section~\ref{discussion}. The supplementary materials contain theoretical proofs.

	\section{Modeling}
	\label{model}
\begin{figure}
		\centering
		\includegraphics[width=70mm]{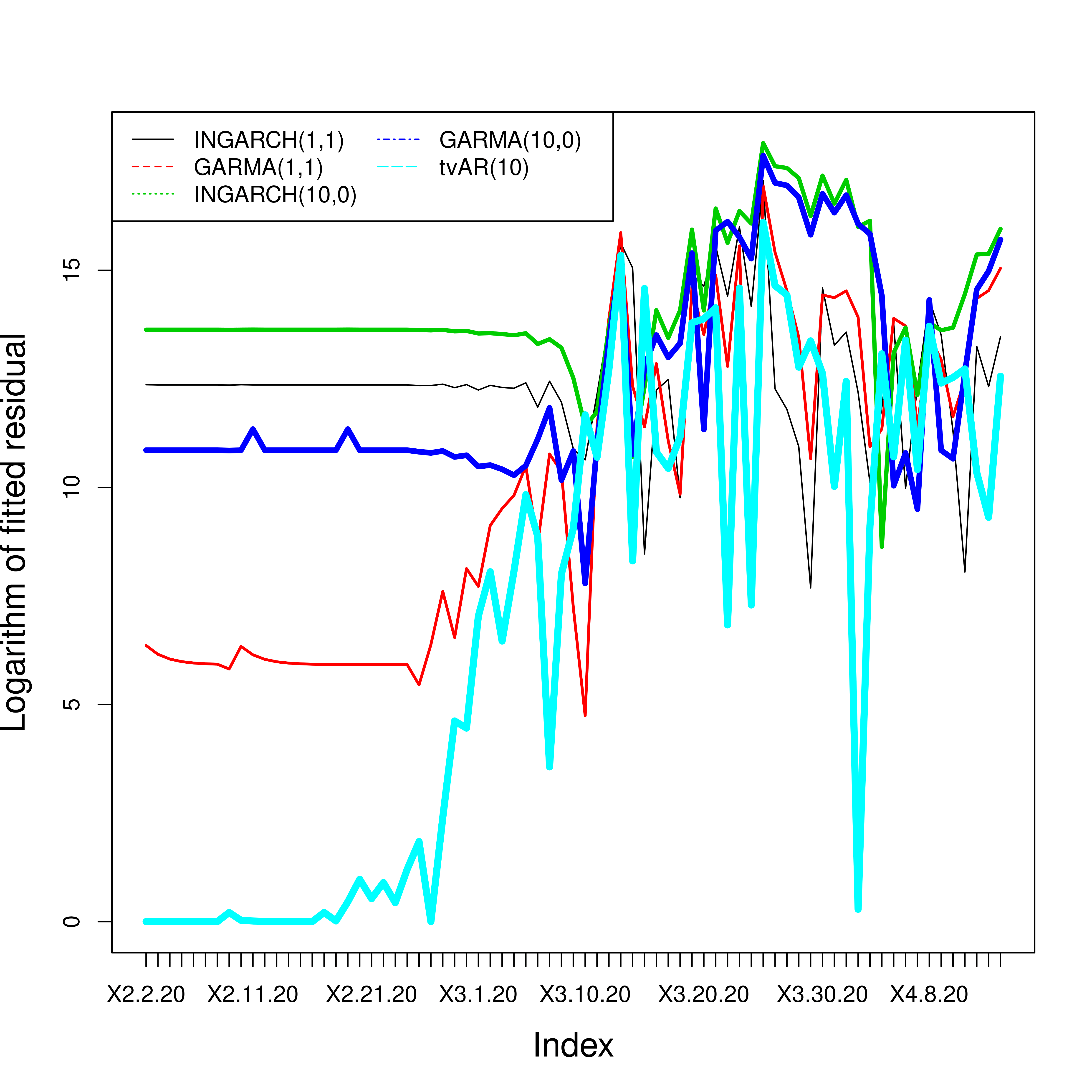}
		\caption{Logarithm of fitted absolute residual for different methods on the COVID-19 spread data from Spain.}
		\label{fig:motiv}
	\end{figure}

	In this paper, our primary focus is on modeling the daily count of newly confirmed cases of COVID-19 in a non-parametric way. Instead of compartmental models from epidemiology, we choose to focus on building a model solely out of the COVID-19 dataset. Figure~\ref{fig:motiv} illustrates the logarithm of the absolute values of fitted residuals for different methods on the COVID-19 spread data from Spain. We find that the time-varying AR model fits the data much better than other routinely used time-constant methods. Given the current knowledge of the incubation period of the virus \citep{lauer2020incubation}, we fit models until lag 10 for the autoregressive part. For the conditional heteroscedastic models, we consider the order as (1,1), which is a standard choice for such models. This motivates us to consider the non-stationary time-series models such as usual tvAR regression. However, we emphasize that these existing methods are suitable only for continuous valued variables and not for count-valued series.
	
	This motivation sets up the stage to discuss Poisson autoregression with time-varying coefficients. Let $\{X_t\}$ be a count-valued time series. In this paper, we consider two different structures for the conditional mean of $\{X_t\}$ given the history of the process. The first modeling framework is in the spirit of time-varying auto-regressive models. Next, we consider another modeling structure in the direction of time-varying generalized autoregressive conditional heteroscedasticity models.
	
	\subsection{Time-varying auto-regressive model for counts}
	The linear Poisson autoregressive model \citep{zeger1988regression, brandt2001linear} is popular in analyzing count valued time series. Due to the assumed non-stationary nature of the data, we propose a time-varying version of this model. The conditional distribution for count-valued time-series $X_t$ given $\sF_{t-1}=\{X_i: i\leq (t-1)\}$ is,
	\begin{align}
	X_t|\sF_{t-1}\sim& \mathrm{Poisson}(\lambda_t) \text{ where } \lambda_t=\mu(t/T) + \sum_{i=1}^p a_i(t/T) X_{t-i}.\label{TVBARC}
	\end{align}
	We call our method time-varying Bayesian Auto Regressive model for Counts (TVBARC). The rescaling of the time-varying parameters to the support [0,1] is usual for in-filled asymptotics. Due to the Poisson link in~\eqref{TVBARC}, both conditional mean and conditional variance depend on the past observations. The conditional expectation of $X_t$ in the above model \eqref{TVBARC} is $\IE(X_t|\sF_{t-1})=\mu(t/T) + \sum_{i=1}^p a_i(t/T) X_{t-i}$, which is positive-valued. Additionally, we impose the following constraints on parameter space for the time-varying parameters, 
	\begin{align}\label{eq:parcondition}
	\mathcal{P}_1=\{\mu, a_i:\mu(x)> 0, 0\leq a_{i}(x)\leq 1, \sup_{x}\sum_{k}a_{k}(x)<1\}.
	\end{align}
	Note that, the conditions imposed (\ref{eq:parcondition}) on the parameters is somewhat motivated from the stationarity conditions for the time-constant versions of this models. This is not uncommon in time-varying AR literature. See \cite{dahlhaussubbarao2006, subbarao2008, sayar2020} for example. Even though the condition on $\mu(\cdot)$ seem restrictive in the light of what we need for invertible time-constant AR(p) process with Gaussian error, it is not unusual when it is used to model variance parameters to ensure positivity; it was unanimously imposed for all the literature mentioned above. Additionally, the above references heavily depend on local stationarity: namely, for every rescaled time  $0<t<1$, they assume the existence of an $\tilde{X}_i$ process which is close to the observed process. One key advantage of our proposal is it is free of any such assumption. Our assumption of only the first moment is also very mild. Moreover, except for a very general linear model discussed in \citep{sayar2020}, to the best of our knowledge, this is the very first analysis of the time-varying parameter for count time-series modeled by Poisson regression. Thus we choose to focus on the methodological development rather than proving the optimality of these conditions. When $p=0$, our proposed model reduces to routinely used nonparametric Poisson regression model as in \cite{shen2015adaptive}.
	
	To proceed with Bayesian computation, we put priors on the unknown functions $\mu(\cdot)$ and $a_i(\cdot)$'s such that they are supported in $\mathcal{P}_1$. The prior distributions on these functions are induced through basis expansions in B-splines with suitable constraints on the coefficients to impose the shape constraints as in $\mathcal{P}$. Detail description of the priors are given below,
	\begin{align}
	\mu(x) =&\sum_{j=1}^{K_1}\exp(\beta_j)B_j(x)\label{prior1}\\
	a_{i}(x)=&\sum_{j=1}^{K_2}\theta_{ij}M_{i}B_j(x), \quad 0\leq\theta_{ij}\leq 1,\\
	M_i=&\frac{\exp(\delta_i)}{\sum_{k=0}^p{\exp(\delta_k)}}, \quad i=1,\ldots,p,\\
	\delta_l\sim&N(0, c_1),\textrm{ for }0\leq l\leq p,\\
	\beta_{j}\sim & N(0, c_2)\textrm{ for } 1\leq j\leq K_1,\\
	\theta_{ij}\sim& U(0,1)\textrm{ for }1\leq i\leq p, 1\leq j\leq K_2\label{prior2}.
	\end{align}
	Here $B_{j}$'s are the B-spline basis functions. The parameters $\delta_{j}$'s are unbounded. 
	
	The prior induced by above construction are $\mathcal{P}$-supported. The verification is very straightforward. In above construction, $\sum_{j=0}^PM_j=1$. Thus $\sum_{j=1}^PM_j\leq 1$. Since $0\leq\theta_{ij}\leq 1$, $\sup_{x}a_i(x)\leq M_i$. Thus $\sup_x\sum_{i=1}^Pa_{i}(x)\leq \sum_{i=1}^PM_i\leq 1$. We have $\sum_{j=1}^PM_j\leq 1$ if and only if $\delta_0=-\infty$, which has probability zero. On the other hand, we also have $\mu(\cdot)\geq 0$ as we have $\exp(\beta_j)\geq 0$. Thus, the induced priors, described in~\eqref{prior1}$-$~\eqref{prior2} are well supported in $\mathcal{P}$.
	
	\subsection{Time-varying generalized autoregressive conditional heteroscedasticity model for counts}
	
	In the previous model, both conditional mean and conditiona variance depend on the past observations. However, \cite{ferland2006integer} proposed integer valued analogue of generalized autoregressive conditional heteroscedasticity model (GARCH) after observing that the variability in number of cases of campylobacterosis
	infections also changes with level. Given the complexity of COVID-19 data, we also introduce the following time-varying version of the integer valued generalized autoregressive conditional heteroscedasticity model (INGARCH) for counts. The conditional distribution for count-valued time-series $X_t$ given $\sF_{t-1}=\{X_i: i\leq (t-1)\}$ and $\sG_{t-1}=\{\lambda_i: i\leq (t-1)\}$ is,
	\begin{align}
	X_t|\sF_{t-1},\sG_{t-1}\sim& \mathrm{Poisson}(\lambda_t) \text{ where } \lambda_t=\mu(t/T) + \sum_{i=1}^p a_i(t/T) X_{t-i}+\sum_{j=1}^q b_j(t/T)\lambda_{t-j}.\label{TVBING}
	\end{align}
	We call our method time-varying Bayesian Integer valued Generalized Auto Regressive Conditional Heteroscedastic (TVBINGARCH) model. We impose following constraints on the parameter space similar to \cite{ferreira2017estimation},
	\begin{align}\label{eq:parconditionG}
	\mathcal{P}_2=\{\mu, a_i:\mu(x)> 0, 0\leq a_{i}(x)\leq 1, 0\leq a_{j}(x)\leq 1, \sup_{x}\sum_{i,j}(a_{i}(x)+b_j(x))<1\}.
	\end{align}
	This constraint ensure a unique solution of the time-varying GARCH process as discussed in \cite{ferreira2017estimation}. Now, we modify the proposed prior from the previous subsection to put prior on the functions $\mu(\cdot)$, $a_i(\cdot)$ and $b_j(\cdot)$ such that they are supported in $\mathcal{P}_2$. Using the B-spline bases, we put following hierarchical prior on the unknown functions,
	\begin{align}
	\mu(x) =&\sum_{j=1}^{K_1}\exp(\beta_j)B_j(x)\label{prior3}\\
	a_{i}(x)=&\sum_{j=1}^{K_2}\theta_{ij}M_{i}B_j(x), \quad 0\leq\theta_{ij}\leq 1,1\leq i\leq p,\\
	b_{k}(x)=&\sum_{j=1}^{K_3}\eta_{kj}M_{k+p}B_j(x), \quad 0\leq\eta_{kj}\leq 1,1\leq k\leq q,\\
	M_i=&\frac{\exp(\delta_i)}{\sum_{k=0}^p{\exp(\delta_k)}}, \quad i=1,\ldots,p+q,\\
	\delta_l\sim&N(0, c_1),\textrm{ for }0\leq l\leq p+q,\\
	\beta_{j}\sim & N(0, c_2)\textrm{ for } 1\leq j\leq K_1,\\
	\theta_{ij}\sim& U(0,1)\textrm{ for }1\leq i\leq p, 1\leq j\leq K_2,\\
	\eta_{kj}\sim& U(0,1)\textrm{ for }1\leq k\leq q, 1\leq j\leq K_3,\\
	\lambda_0\sim&\textrm{Inverse-Gamma}(d_1,d_1)\label{prior4}.
	\end{align}
	Similar calculations from previous subsection also shows that the above hierarchical prior in~\eqref{prior3} to~\eqref{prior4} is well-supported in $\mathcal{P}_2$. We primarily focus on the special case where $p=1,q=1$. 

	\section{Posterior computation}
	\label{comp}
	In this section, we discuss Markov Chain Monte Carlo (MCMC) sampling method for posterior computation. Our proposed sampling is dependent on gradient based Hamiltonian Monte Carlo (HMC) sampling algorithm \citep{neal2011mcmc}. Hence, we show the gradient computations of the likelihood with respect to different parameters for TVBARC$(p)$ and TVBINGARCH$(p,q)$ in following two subsections. 
	\subsection{TVAR structure}
	The complete likelihood $L$ of the propose Bayesian method in~\eqref{TVBARC} is given by
	\begin{align*}
	L_1&\propto \exp\bigg(\sum_{t=p}^T \big[-\{\mu(t/T) + \sum_{i=1}^p a_i(t/T) X_{t-i}\big\} + X_t\log \big\{\mu(t/T) \\
	&\quad+ \sum_{i=1}^p a_i(t/T) X_{t-i}\}\big] - \sum_{j=1}^{K_1} \beta_j^2/(2c_2) - \sum_{l=0}^p \delta_l^2/(2c_1)\bigg){\mathbf 1}_{0\leq\theta_{ij}\leq 1},
	\end{align*}
	where $\mu(x) =\sum_{j=1}^{K_1}\exp(\beta_j)B_j(x), a_{i}(x)=\sum_{j=1}^{K_2}\theta_{ij}M_{i}B_j(x)$ and $M_j=\frac{\exp(\delta_j)}{\sum_{k=0}^j{\exp(\delta_k)}}$. We develop efficient MCMC algorithm to sample the parameter $\beta,\theta$ and $\delta$ from the above likelihood. The derivatives of above likelihood with respect to the parameters are easily computable. This helps us to develop an efficient gradient-based MCMC algorithm to sample these parameters. We calculate the gradients of negative log-likelihood $(-\log L_1)$ with respect to the parameters $\beta$, $\theta$ and $\delta$. The gradients are given below,
	\begin{align*}
	&-\frac{d\log L_1}{\beta_j}=\exp(\beta_j)\bigg(1-\sum_t  \frac{B_j(t/T)X_t}{(\mu(t/T)+\sum_{j}a_{j}(t/T)X_{t-j})}\bigg) + \beta_j/c_2,\\
	&-\frac{d\log L_1}{\theta_{ij}}=M_{j}\bigg(1-\sum_t \frac{B_{j}(t/T)X_t}{(\mu(t/T)+\sum_{j}a_{j}(t/T)X_{t-j})}\bigg)&,\\
	&-\frac{d\log L_1}{\delta_j}=\delta_j/c_1+\nonumber\\&\quad\sum_k (M_j{\mathbf 1}_{\{j=k\}}-M_jM_k)\sum_i\theta_{ij}B_j(x)\bigg(1-\sum_t \frac{B_{j}(t/T)X_{t-j}}{(\mu(t/T)+\sum_{j}a_{j}(t/T)X_{t-j})}\bigg),
	\end{align*}
	where ${\mathbf 1}_{\{j=k\}}$ stands for the indicator function which takes the value one when $j=k$. 
	
	\subsection{TVBINGARCH structure}
	The complete likelihood $L_2$ of the propose Bayesian method of~\eqref{TVBING} is given by
	\begin{align*}
	L_2&\propto \exp\bigg(\sum_{t=p}^T \big[-\{\mu(t/T) + \sum_{i=1}^p a_i(t/T) X_{t-i}+\sum_{i=1}^q b_i(t/T) \lambda_{t-i}\big\} + X_t\log \big\{\mu(t/T) \\
	&\quad+ \sum_{i=1}^p a_i(t/T) X_{t-i}+\sum_{i=1}^q b_i(t/T) \lambda_{t-i}\}\big] - \sum_{j=1}^{K_1} \beta_j^2/(2c_2) - \sum_{l=0}^p \delta_l^2/(2c_1)\\
	&\quad-(d_1+1)\log\lambda_0 -d_1/\lambda_0\bigg){\mathbf 1}_{0\leq\theta_{ij},\eta_{ij}\leq 1},
	\end{align*}
	We calculate the gradients of negative log-likelihood $(-\log L_2)$ with respect to the parameters $\beta$, $\theta$, $\eta$ and $\delta$. The gradients are given below,
	\begin{align*}
	&-\frac{d\log L_2}{\beta_j}=\exp(\beta_j)\bigg(1-\sum_t  \frac{B_j(t/T)X_{t-j}}{(\mu(t/T)+\sum_{j}a_{j}(t/T)X_{t-j})+\sum_{k}b_{k}(t/T)\lambda_{t-k})}\bigg) + \beta_j/c_2,\\
	&-\frac{d\log L_2}{\theta_{ij}}=M_{j}\bigg(1-\sum_t \frac{B_{j}(t/T)X_{t-j}}{(\mu(t/T)+\sum_{j}a_{j}(t/T)X_{t-j})+\sum_{k}b_{k}(t/T)\lambda_{t-k})}\bigg)&,\\
	&-\frac{d\log L_2}{\eta_{kj}}=M_{j}\bigg(1-\sum_t \frac{B_{j}(t/T)\lambda_{t-j}}{(\mu(t/T)+\sum_{j}a_{j}(t/T)X_{t-j})+\sum_{k}b_{k}(t/T)\lambda_{t-k})}\bigg)&,\\
	&-\frac{d\log L_2}{\delta_j}=\delta_j/c_1+\sum_k (M_j{\mathbf 1}_{\{j=k\}}-M_jM_k)\times \nonumber\\&\Bigg[\quad\sum_{i\leq p}\theta_{ij}B_j(x)\bigg(1-\sum_t \frac{B_{j}(t/T)X_{t-j}}{(\mu(t/T)+\sum_{j}a_{j}(t/T)X_{t-j})+\sum_{k}b_{k}(t/T)\lambda_{t-k})}\bigg){\mathbf 1}_{\{j\leq p\}}+\nonumber\\
	&\quad\sum_{1\leq k\leq q}\eta_{kj}B_j(x)\bigg(1-\sum_t \frac{B_{j}(t/T)\lambda_t}{(\mu(t/T)+\sum_{j}a_{j}(t/T)X_{t-j})+\sum_{k}b_{k}(t/T)\lambda_{t-k})}\bigg){\mathbf 1}_{\{j > p\}}\Bigg].
	\end{align*}
	While fitting TVBINGARCH(p,q), we assume for any $t<0$ $X_t=0,\lambda_t=0$. Thus, we need to additionally estimate the parameter $\lambda_0$. The derivative of the likelihood concerning $\lambda_0$ is calculated numerically using the {\tt jacobian} function from R package {\tt pracma}. Hence, it is sampled using the HMC algorithm too.

	As the parameter spaces of $\theta_{ij}$'s and $\eta_{kj}$'s have bounded support, we map any Metropolis candidate, falling outside of the parameter space back to the nearest boundary point of the parameter space. The number of leapfrog steps is kept fixed at 30, however, the step size parameter is tuned to maintain an acceptance rate within the range of 0.6 to 0.8. The step length is reduced if the acceptance rate is less than 0.6 and increased it if the rate is more than 0.8. This adjustment is done automatically after every 100 iterations. Due to the increasing complexity of the parameter space in TVBINGARCH, we propose to update all the parameters involved in $a_{i}(\cdot)$'s and $b_k(\cdot)$'s together.
	
	\section{Simulation studies}
	\label{sim}
	In this section, we study the performance of our proposed Bayesian method in capturing the true coefficient functions. We explore both TVBARC and TVBINGARCH with some competing models. It is important to note that, this is to the best of our knowledge first work in Poisson autoregression with a time-varying link. Thus, we compare our method with the existing time-series models with time-constant coefficients for count data and time-varying AR with Gaussian error. We also examine the estimation accuracy of the coefficient functions to the truth. 
	
	The hyperparameters $c_1$ and $c_2$ of the normal prior are all set 100, which makes the prior weakly informative. The hyperparmaters for Inverse-Gamma prior $d_1=0.1$, which is also weakly informative. We consider 6 equidistant knots for the B-splines. We collect 10000 MCMC samples and consider the last 5000 as post burn-in samples for inferences. In absence of any alternative method for time-varying AR$(p)$ model of count-valued data, we shall compare the estimated functions with the true functions in terms of the posterior estimates of functions along with its 95\% pointwise credible bands. The credible bands are calculated from the MCMC samples at each point $t=1/T,2/T,\ldots, 1$. We also compare different competing methods in terms of average MSE (AMSE) score using the INGARCH method of {\tt tsglm} of R package {\tt tscount}, GARMA using {\tt tscount} as well, {\tt tvAR} and our proposed Bayesian methods.  We define AMSE as $\frac{1}{T}\sum_t(X_t-\hat{\lambda}_t)^2$.

	\subsection{Case 1: TVBARC structure}
	\label{TVARsec}
	
	Here, we consider two model settings $p=1;X_t\sim\textrm{Poisson}(\mu(t/T)+a_1(t/T)X_{t-1})$ and $p=2;X_t\sim\textrm{Poisson}(\mu(t/T)+a_1(t/T)X_{t-1} + a_2(t/T)X_{t-2})$ for $t=1,\ldots,T$. Three different choices for $T$ have been considered, $T=100,500$ and $1000$. The true functions are,
	\begin{align*}
	\mu_0(x)=&10\exp\big(-(x-0.5)^2/0.1\big),\\
	a_{10}(x)=&0.3(x-1)^2+0.1,\\
	a_{02}(x)=&0.4x^2+0.1.
	\end{align*}
	
\begin{figure}[htbp]
		\centering
		\subfigure[$\mu()$]{\label{fig:c.1}\includegraphics[width=50mm]{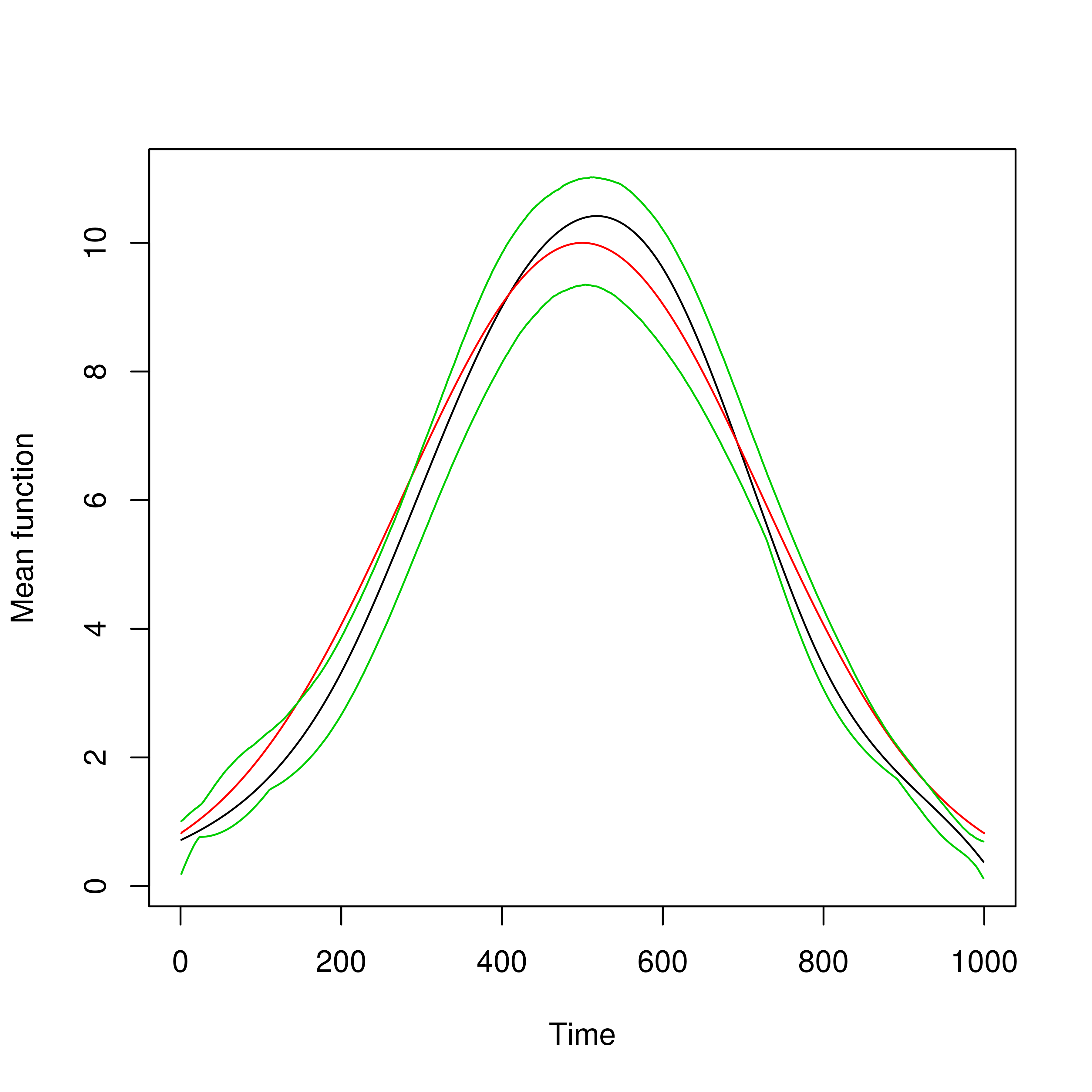}}
		\subfigure[$a_1()$]{\label{fig:c.2}\includegraphics[width=50mm]{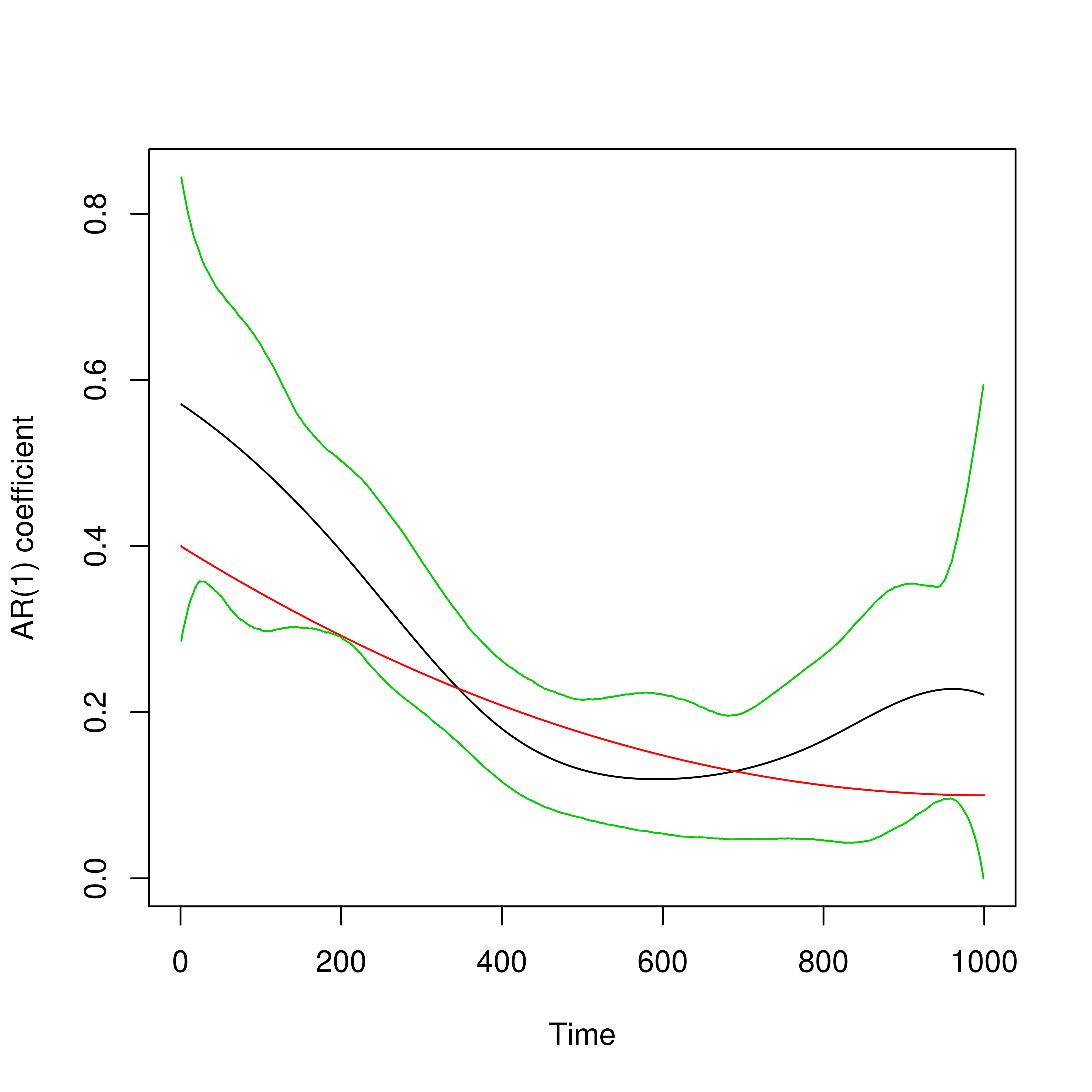}}
		\caption{Estimated mean function in 1st column and estimated AR(1) coefficient function in the 2nd column for the case $p=1$ and sample size 1000. Red is the true function, black is the estimated curve along with the 95\% pointwise credible bands in green.} 
		\label{AR1}
	\end{figure}

\begin{figure}[htbp]
		\centering
		\includegraphics[width=120mm]{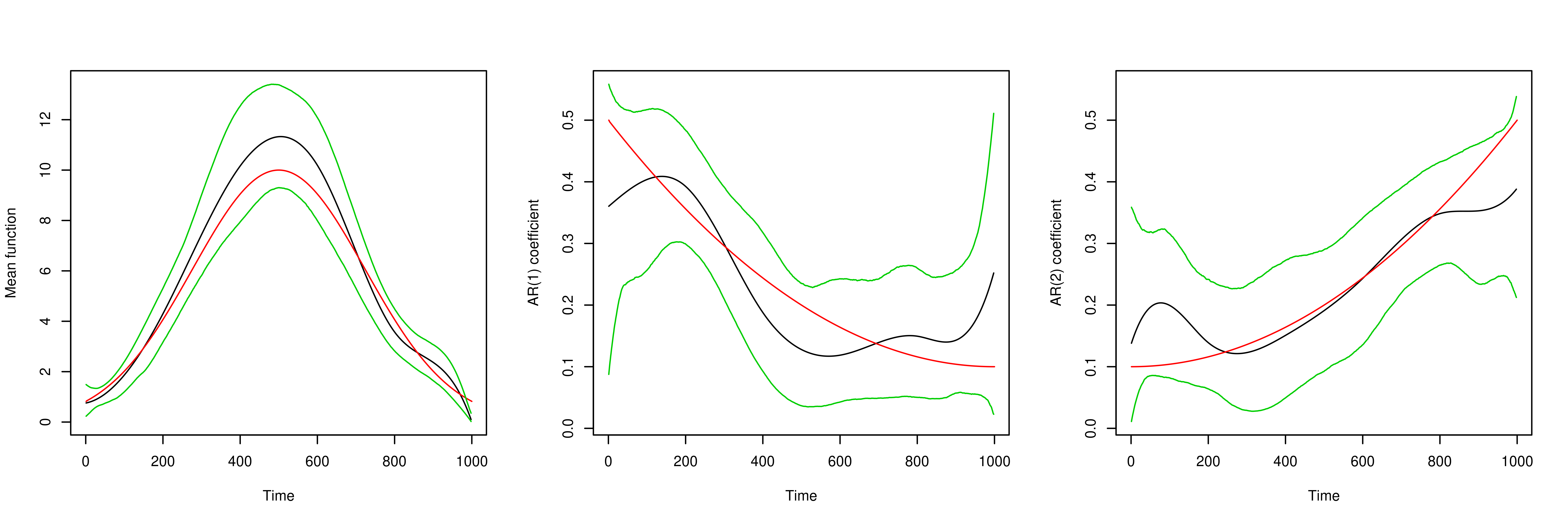}
		\caption{Estimated coefficient functions for the simulation case $p=2$ and sample size 1000. Red is the true function, black is the estimated curve along with the 95\% pointwise credible bands in green.} 
		\label{AR2}
	\end{figure}

	We compare the estimated functions with the truth for sample size 1000 in Figures~\ref{AR1} and Figure~\ref{AR2} for the models $p=1$ and $p=2$ respectively. Tables~\ref{AMSEAR1} and~\ref{AMSEAR2} illustrate the performance of our method with respect to other competing methods.
	
	\begin{table}[ht]
		\centering
		\caption{AMSE comparison for different sample sizes across different methods when the true model is~\eqref{TVBARC} with $p=1$.}
		\begin{tabular}{rrrrr}
			\hline
			& INGARCH(1,0) & GARMA(1,0) & TVAR(1) & TVBARC(1) \\ 
			\hline
			$T=100$ & 11.60 & 11.18 & 11.41 & \textbf{8.65} \\  
			$T=500$ & 11.35 & 11.04 & 11.24 & \textbf{8.12} \\ 
			$T=1000$ & 11.05 & 10.73 & 10.94 & \textbf{7.02} \\ 
			\hline
		\end{tabular}
		\label{AMSEAR1}
	\end{table}
	
	\begin{table}[ht]
		\centering
		\caption{AMSE comparison for different sample sizes across different methods when the true model is~\eqref{TVBARC} with $p=2$.}
		\begin{tabular}{rrrrr}
			\hline
			& INGARCH(2,0) & GARMA(2,0) & TVAR(2) & TVBARC(2) \\ 
			\hline
			$T=100$ & 18.02 & 17.28 & 13.04 & \textbf{11.01} \\ 
			$T=500$ & 16.42 & 15.86 & 12.61 & \textbf{10.79} \\ 
			$T=1000$ & 15.79 & 15.25 & 12.75 & \textbf{10.61} \\ 
			\hline
		\end{tabular}
		\label{AMSEAR2}
	\end{table}
	
	\subsection{Case 2: TVBINGARCH structure}
	\label{TVBINGARCHsec}
	For the tvINGARCH case, we only consider one simulation settings $p=1, q=1; X_t\sim\textrm{Poisson}(\mu(t/T)+a_1(t/T)X_{t-1} +b_1(t/T)\lambda_{t-1})$. Two different choices for $T$ have been considered, $T=100$ and $200$,
	\begin{align*}
	\mu_0(x)=&25\exp\big(-(x-0.5)^2/0.1\big),\\
	a_1(x) =& 0.4x^2+0.1,\\
	b_1(x) =& 0.1\sin(\pi x)+0.2
	\end{align*}
	
\begin{figure}
		\centering
		\includegraphics[width=100mm]{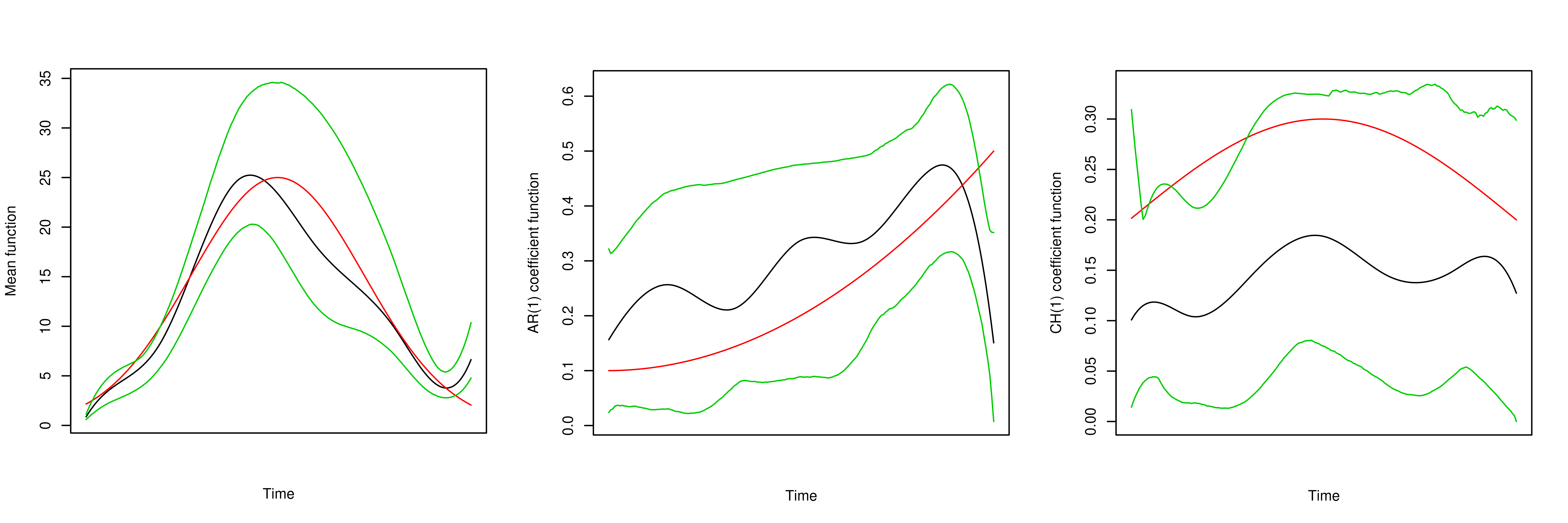}
		\caption{Estimated coefficient functions for the TVBINGARCH(1,1) and sample size 200. Red is the true function, black is the estimated curve along with the 95\% pointwise credible bands in green.}
		\label{TVBINGARCH}
	\end{figure}

	Figures~\ref{TVBINGARCH} compares the estimated functions with the truth for sample size 200 for the model in~\eqref{TVBING} with $p=1,q=1$. The performance of our method is compared to other competing methods in Tables~\ref{TVINGARtab}.
	
	\begin{table}[ht]
		\caption{Average MSE comparison for different sample sizes across different methods when the true model is~\eqref{TVBING} with $p=1, q=1$.}
		\centering
		\begin{tabular}{rrrrr}
			\hline
			& INGARCH(1,1) & GARMA(1,1) & tvAR(1) & TVBINGARCH(1,1) \\ 
			\hline
			$T=100$ & 41.95 & 37.82 & 48.19 & \textbf{30.24} \\ 
			$T=200$ & 37.43 & 36.07 & 42.36 & \textbf{27.84} \\ 
			\hline
		\end{tabular}
		\label{TVINGARtab}
	\end{table}
	
	Figure~\ref{AR1} to~\ref{TVBINGARCH} shows that our proposed Bayesian method captures the true functions quite well for both of the two simulation experiments. We find that the estimation accuracy improves as the sample size increases. As the sample size grows, the 95\% credible bands are also getting tighter, implying lower uncertainty in estimation. This gives empirical evidence in favor of the estimation consistency Even though the credible intervals form a very useful tool to build pointwise inference for the time-trajectory of these coefficient functions, we do not report the coverage probability here. The average mean square error (AMSE) is always the lowest for our method in Tables~\ref{AMSEAR1}, \ref{AMSEAR2} and~\ref{TVINGARtab}. For the Poisson distribution, mean and variance are the same. Since $\mu_0(x)$ is around 10 for simulation case~\ref{TVARsec}, the optimal AMSE will be  around 10, which is achieved by our method. Similarly for simulation case~\ref{TVBINGARCHsec}, the optimal AMSE is expected to be around 25.
	
	\section{COVID-19 data application}
	\label{application}
	We collect the data of new affected cases for every day from 23rd January to 14th April from an open-source platform \url{ {https://www.kaggle.com/sudalairajkumar/novel-corona-virus-2019-dataset}}. Table~\ref{tab:count} provide a summary of total affected cases along with the number of recovered and deceased for three most affected countries along with Hubei, New York City (NYC) and Seoul, South Korea. Seoul is selected in the analysis for its unique approach to curbing the outbreak by implementing an aggressive testing strategy. Hubei, Italy and NYC enforced region-wide strict lockdown on January 23rd, March 10th, and March 20th respectively. The US, in general, has implemented selective lockdowns across the country.
	South Korea took an alternative measure to track the movements of all affected individuals and to conduct tests for COVID-19 for as many countrymen as possible. 
	
	We fit the TVBARC model in~\eqref{TVBARC} with $p=10$ and the TVINGARCH model in~\eqref{TVBING} with $p=1, q=1$ for the selected set of countries. The hyperparameters are the same as in the Section~\ref{sim}. We collect 5000 post-burn samples for inference after burn-in 5000 MCMC samples. We calculate derivatives of the estimated functions using derivatives of B-splines \citep{de2001practical}. Note that the confidence bands around the estimated curves provide an uncertainty quantification and offer us to objectively decide on statistically testing certain time-trends.

	We compile square-root average MSE (AMSE) scores for different methods in  Table~\ref{AMSEreal}. One can see all the time-varying methods are doing exceptionally better compared to the time-constant methods. This is not surprising since the spread distribution shows significant time-nonstationarity. Interestingly, one of the two Bayesian methods we proposed in this paper stands out as the best fit. We analyzed three countries, two cities and the earliest epicenter Hubei separately in our analysis. The mechanism of spread in these regions behave a little differently due to population density, government interference, travel, etc. However, some patterns are very eminently similar to all these regions.

	\begin{table}[ht]
		\centering
		\caption{The total number of affected, recovered and dead cases for the selected geographical regions.}\label{tab:count}
		\begin{tabular}{rrrr}
			\hline
			& Total cases & Recovered & Death \\ 
			\hline
			US & 607670 & 47763 & 25787 \\ 
			Spain & 172541 & 67504 & 18056 \\ 
			Italy & 162488 & 37130 & 21067 \\ 
			Hubei, China & 67803 & 64363 & 3221 \\ 
			Seoul, South Korea & 10564 & 7534 & 222 \\ 
			NYC, USA & 110465 & $-$ & 7349 \\ 
			\hline
		\end{tabular}
	\end{table}
	
	\begin{table}[ht]
		\centering
		\caption{Comparison of squre-root-AMSE for different methods across the selected locations.}
		\begin{tabular}{rrrrrrr}
			\hline
			& US & Spain & Italy & Hubei & NYC & Seoul\\ 
			\hline
			TVBARC(10)  & \textbf{1092.97} & 621.08 & \textbf{496.52} & 1551.21  & \textbf{363.72} & 75.02\\ 
			TVBINGARCH(1,1)  & 3181.41 & \textbf{608.00} & 650.81 & \textbf{1484.96}  & 387.93 & \textbf{67.62} \\ 
			INGARCH(1,1) & 6673.87 & 1048.19 & 1118.87 & 1711.48 & 552.82 & 83.68 \\ 
			GARMA(1,1) & 2569.93 & 1087.00 & 1019.96 & 1782.14 & 607.39 & 88.76 \\ 
			INGARCH(10,0) & 9740.67 & 2292.91 & 1675.16 & 1995.08 & 1480.41 & 185.89 \\ 
			GARMA(10,0) & 8613.32 & 2253.17 & 1672.72 & 1834.75 & 1498.86 & 185.95 \\ 
			tvAR(10) & 1161.02 & 661.10 & 581.03 & 1724.03 & 456.66 & 81.13 \\ 
			\hline
		\end{tabular}
		\label{AMSEreal}
	\end{table}
	
\begin{figure}[htbp]
		\centering
		\subfigure[USA-$\mu(\cdot)$ function]{\label{fig:a.1}\includegraphics[width=60mm]{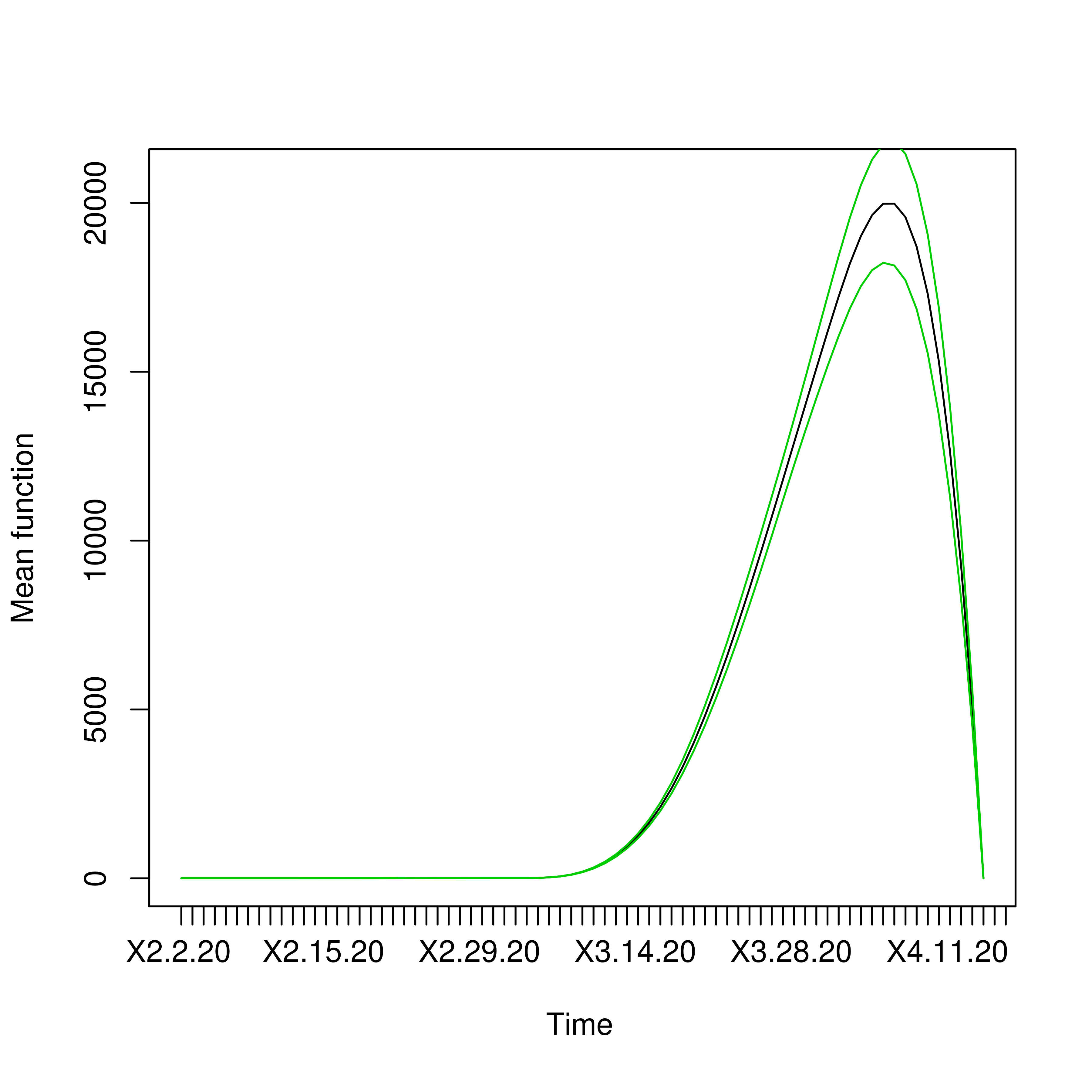}}
		\subfigure[USA-$a(\cdot)$ functions]{\label{fig:a.2}\includegraphics[width=60mm]{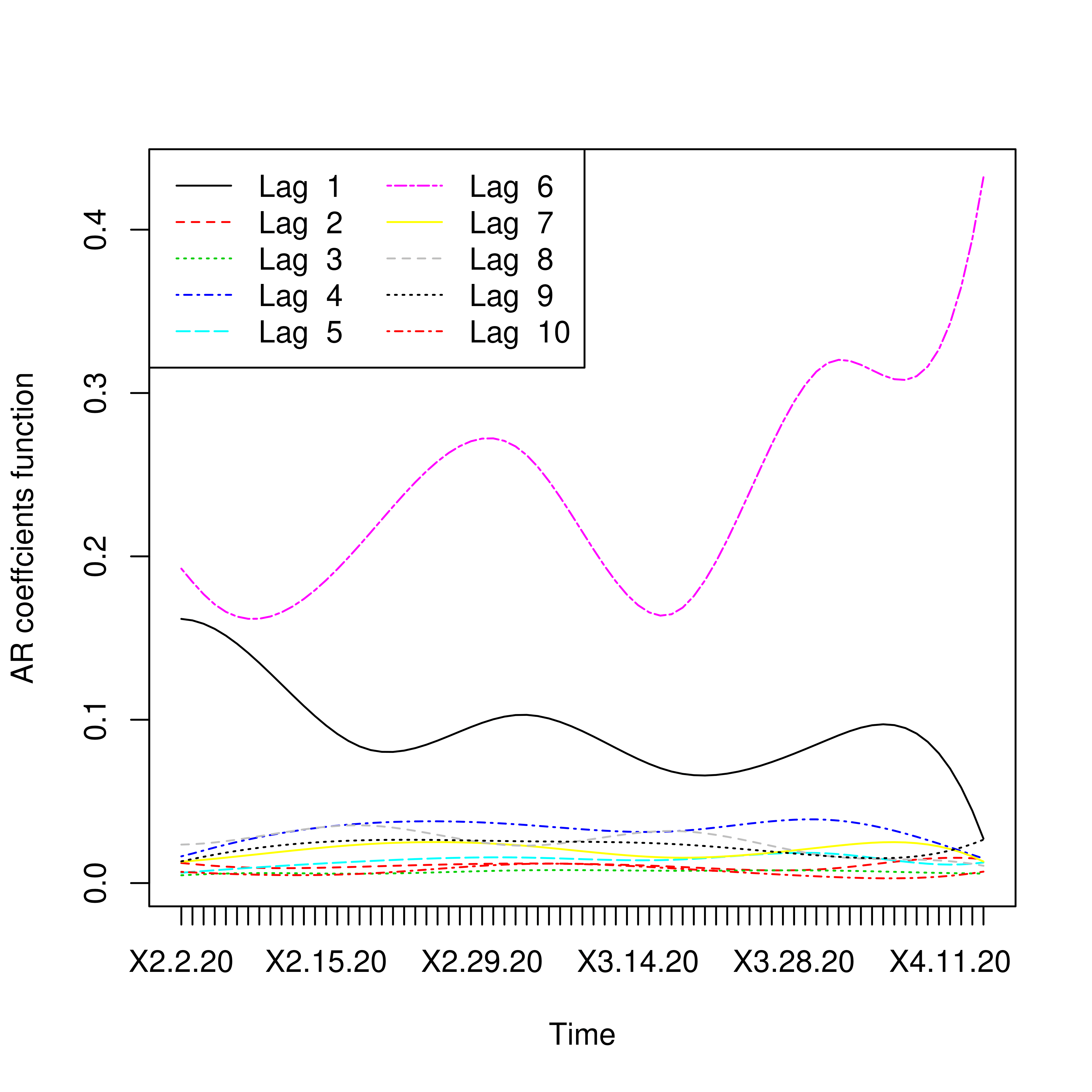}}
		\caption{Estimated mean functions in 1st column and estimated AR coefficient functions in the 2nd column for USA. Black is the estimated curve along with the 95\% pointwise credible bands in green for the mean function.} 
		\label{fig:US}
	\end{figure}

\begin{figure}[htbp]
		\centering
		\subfigure[Italy-$\mu(\cdot)$ function]{\label{fig:a.1}\includegraphics[width=60mm]{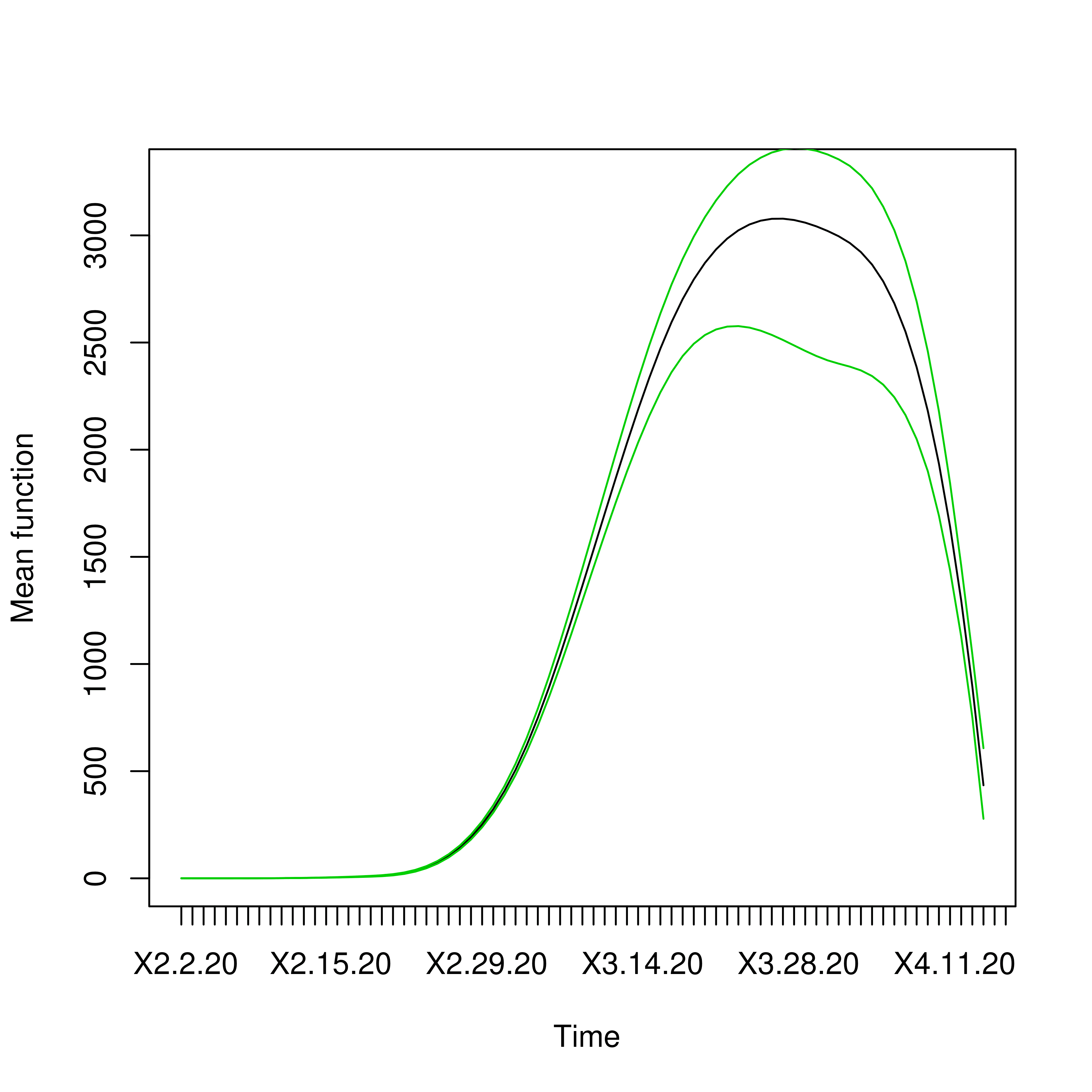}}
		\subfigure[Italy-$a(\cdot)$ functions]{\label{fig:a.2}\includegraphics[width=60mm]{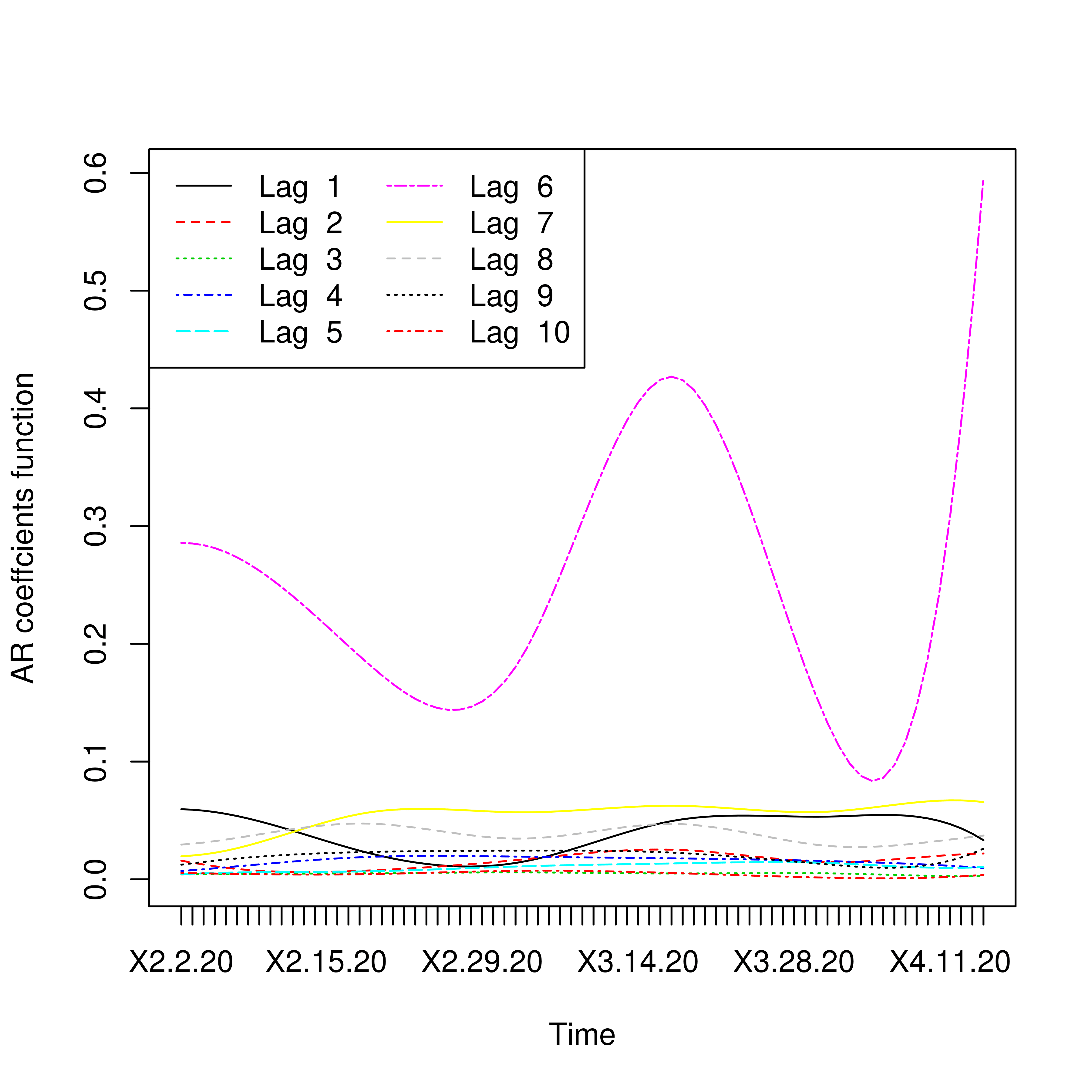}}
		\caption{Estimated mean functions in 1st column and estimated AR coefficient functions in the 2nd column for Italy. Black is the estimated curve along with the 95\% pointwise credible bands in green for the mean function.} 
		\label{fig:Italy}
	\end{figure}

\begin{figure}[htbp]
		\centering
		\subfigure[Spain-$\mu(\cdot)$ function]{\label{fig:a.1}\includegraphics[width=60mm]{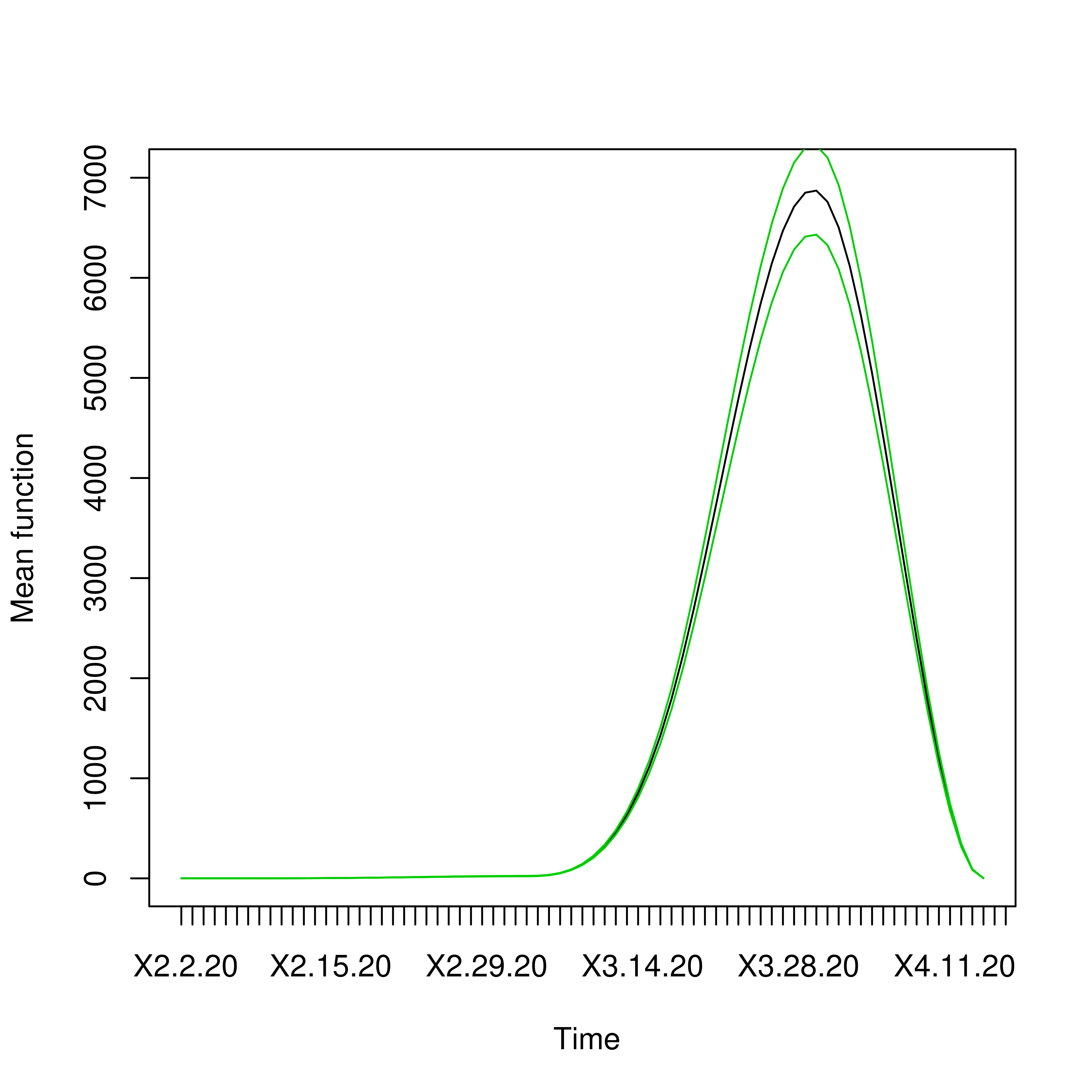}}
		\subfigure[Spain-$a(\cdot)$ functions]{\label{fig:a.2}\includegraphics[width=60mm]{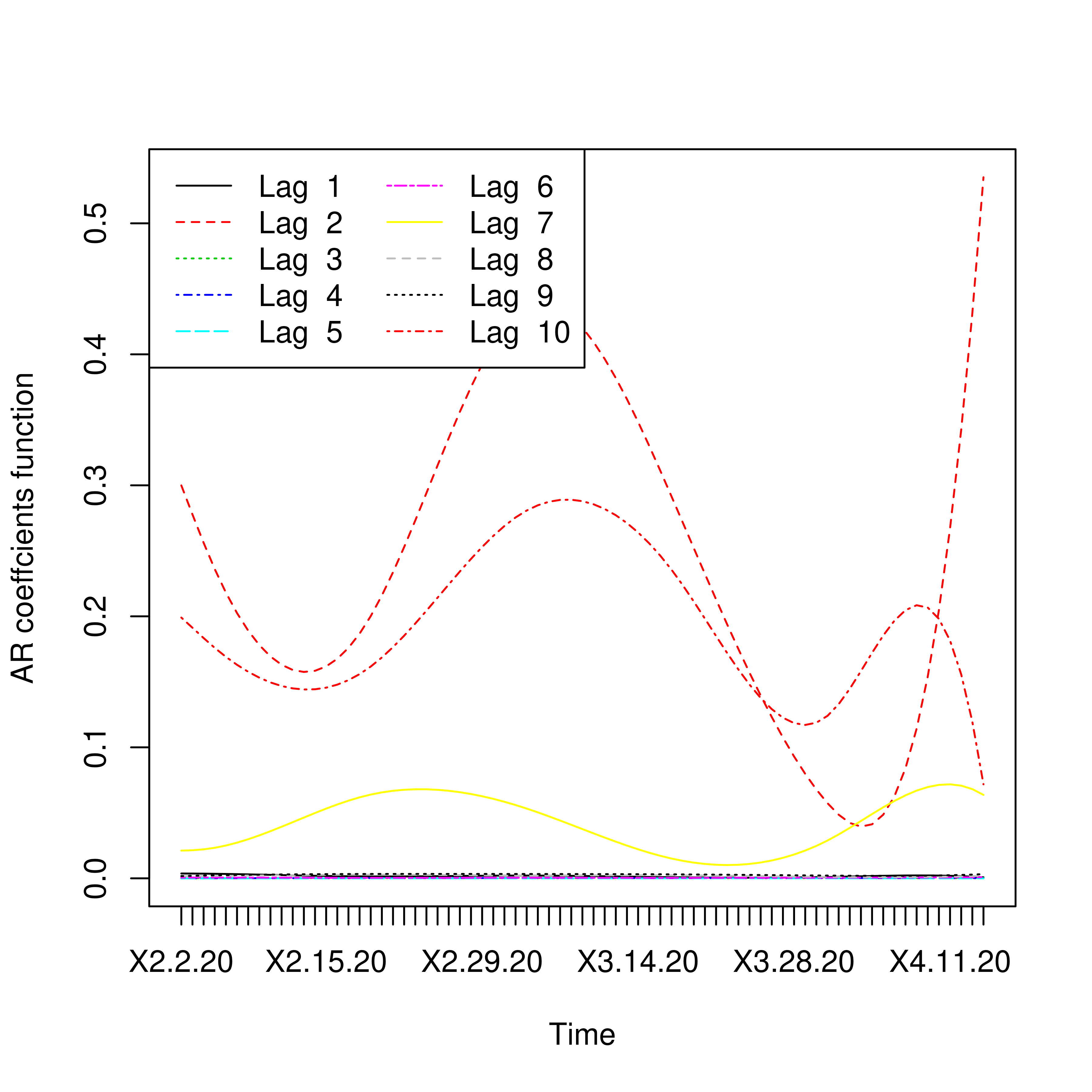}}
		\caption{Estimated mean functions in 1st column and estimated AR coefficient functions in the 2nd column for Spain. Black is the estimated curve along with the 95\% pointwise credible bands in green for the mean function.} 
		\label{fig:Spain}
	\end{figure}

\begin{figure}[htbp]
		\centering
		\subfigure[Hubei, China-$\mu(\cdot)$ function]{\label{fig:a.1}\includegraphics[width=60mm]{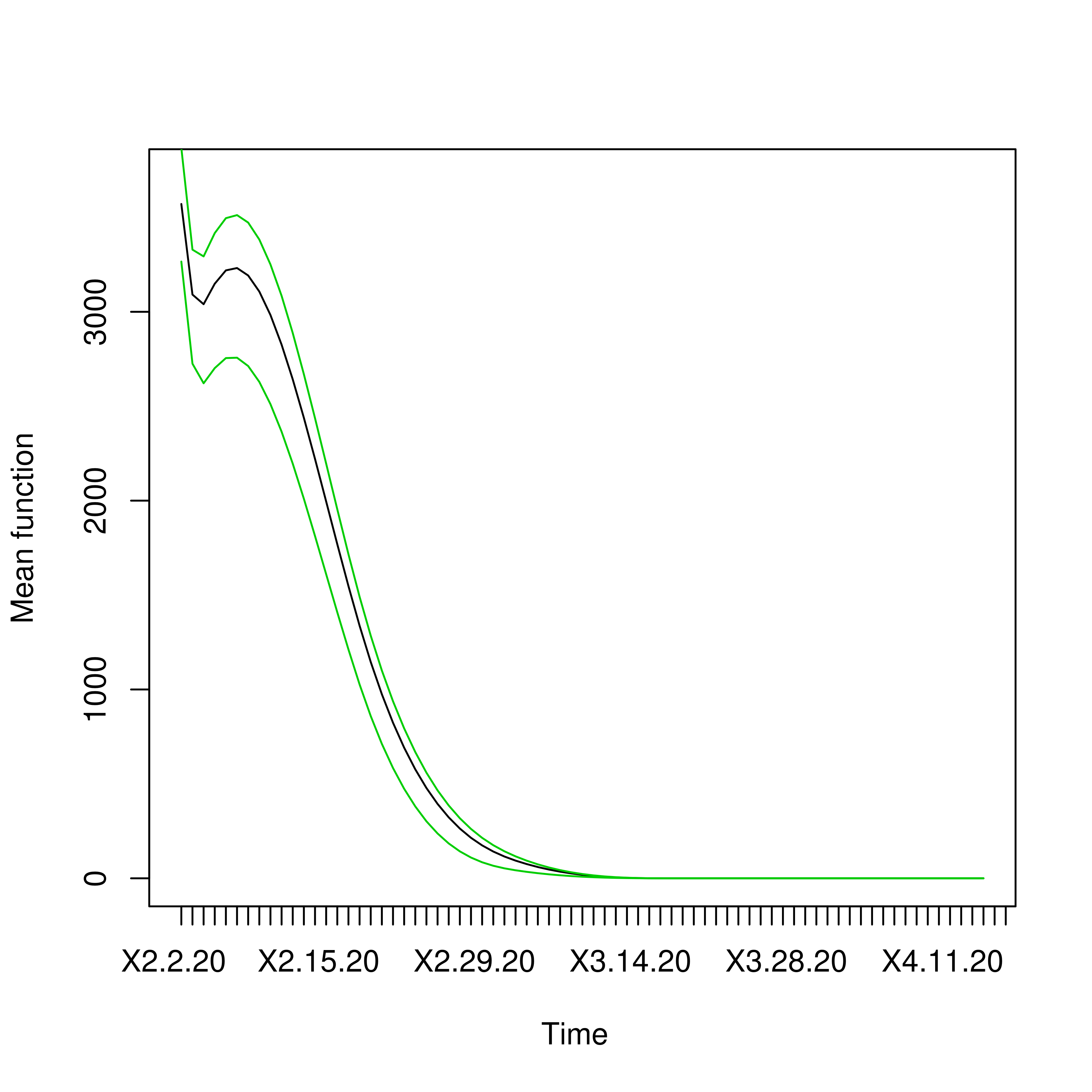}}
		\subfigure[Hubei, China-$a(\cdot)$ functions]{\label{fig:a.2}\includegraphics[width=60mm]{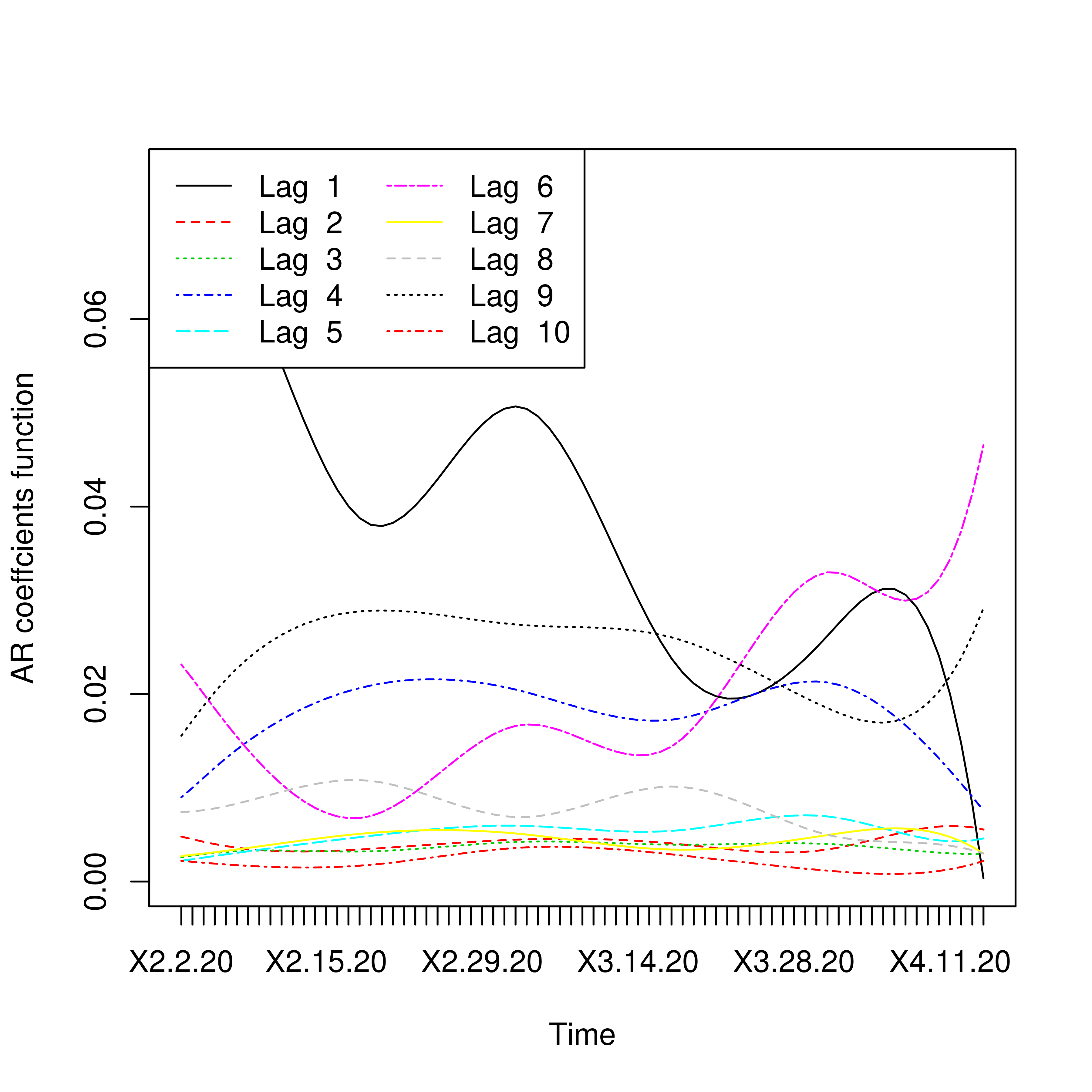}}
		\caption{Estimated mean functions in 1st column and estimated AR coefficient functions in the 2nd column for Hubei, China. Black is the estimated curve along with the 95\% pointwise credible bands in green for the mean function.} 
		\label{fig:hubei}
	\end{figure}

\begin{figure}[htbp]
		\centering
		\subfigure[NYC-$\mu(\cdot)$ function]{\label{fig:a.1}\includegraphics[width=60mm]{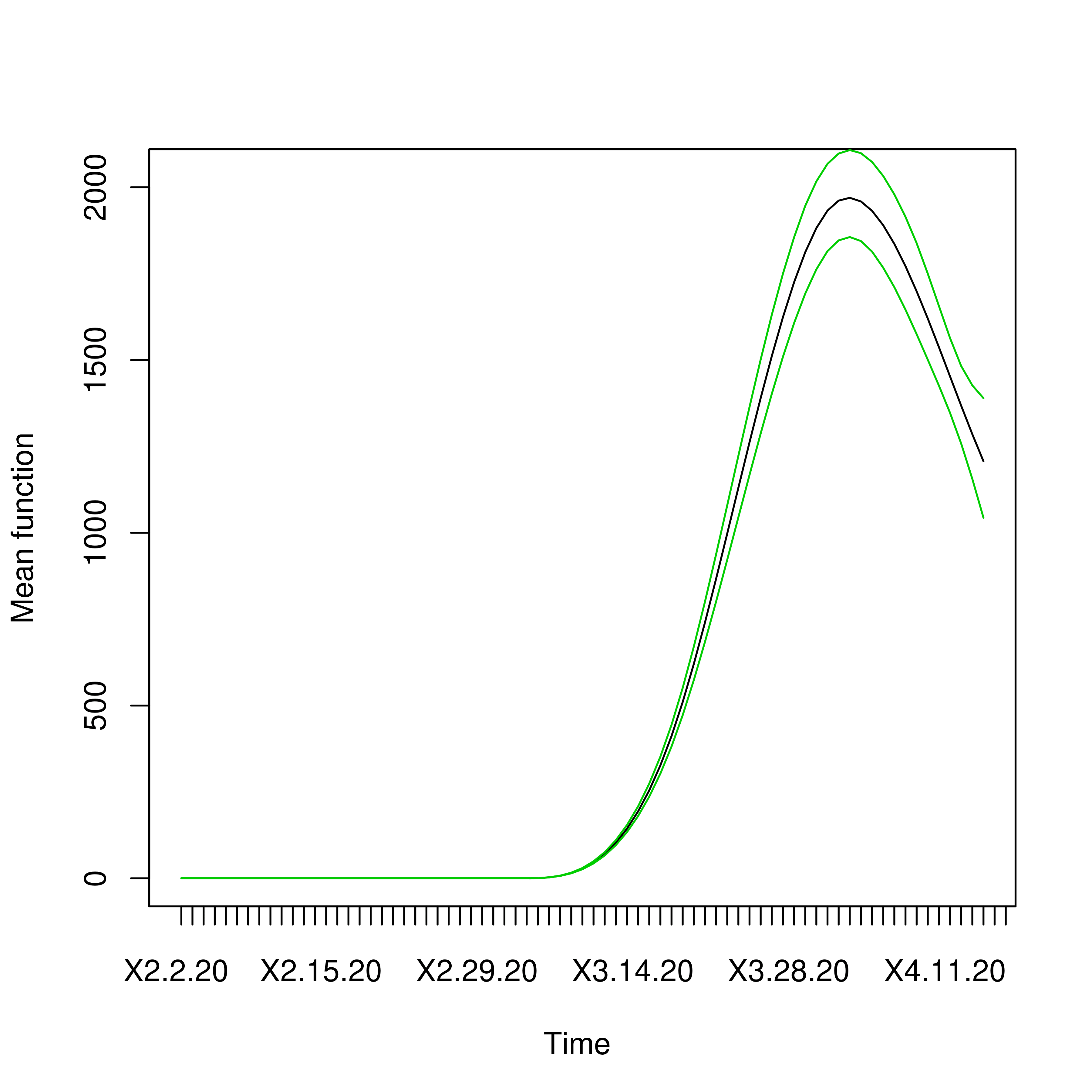}}
		\subfigure[NYC-$a(\cdot)$ functions]{\label{fig:a.2}\includegraphics[width=60mm]{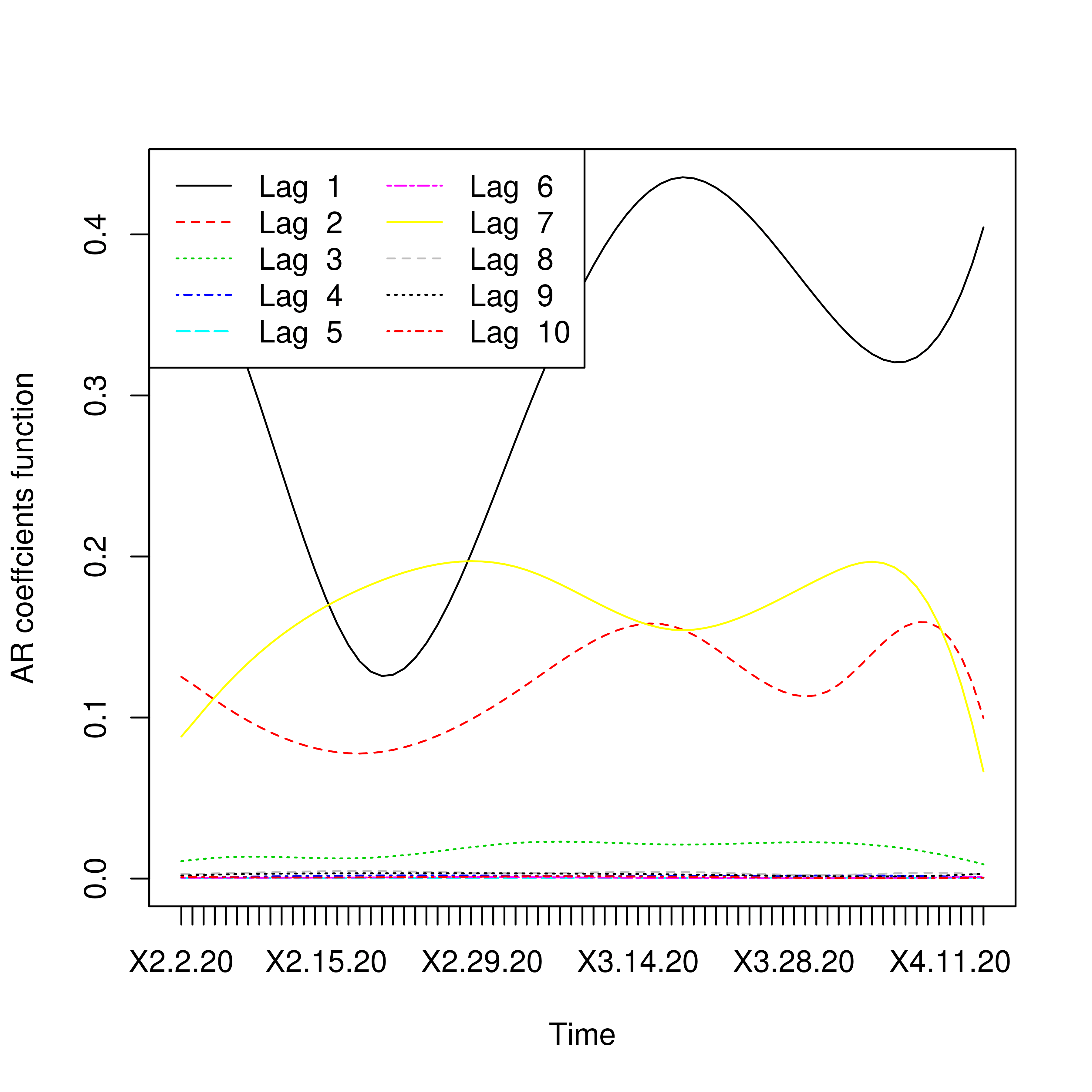}}
		\caption{Estimated mean functions in 1st column and estimated AR coefficient functions in the 2nd column for Hubei, China. Black is the estimated curve along with the 95\% pointwise credible bands in green for the mean function.} 
		\label{fig:NYC}
	\end{figure}

\begin{figure}[htbp]
		\centering
		\subfigure[Seoul-$\mu(\cdot)$ function]{\label{fig:a.1}\includegraphics[width=60mm]{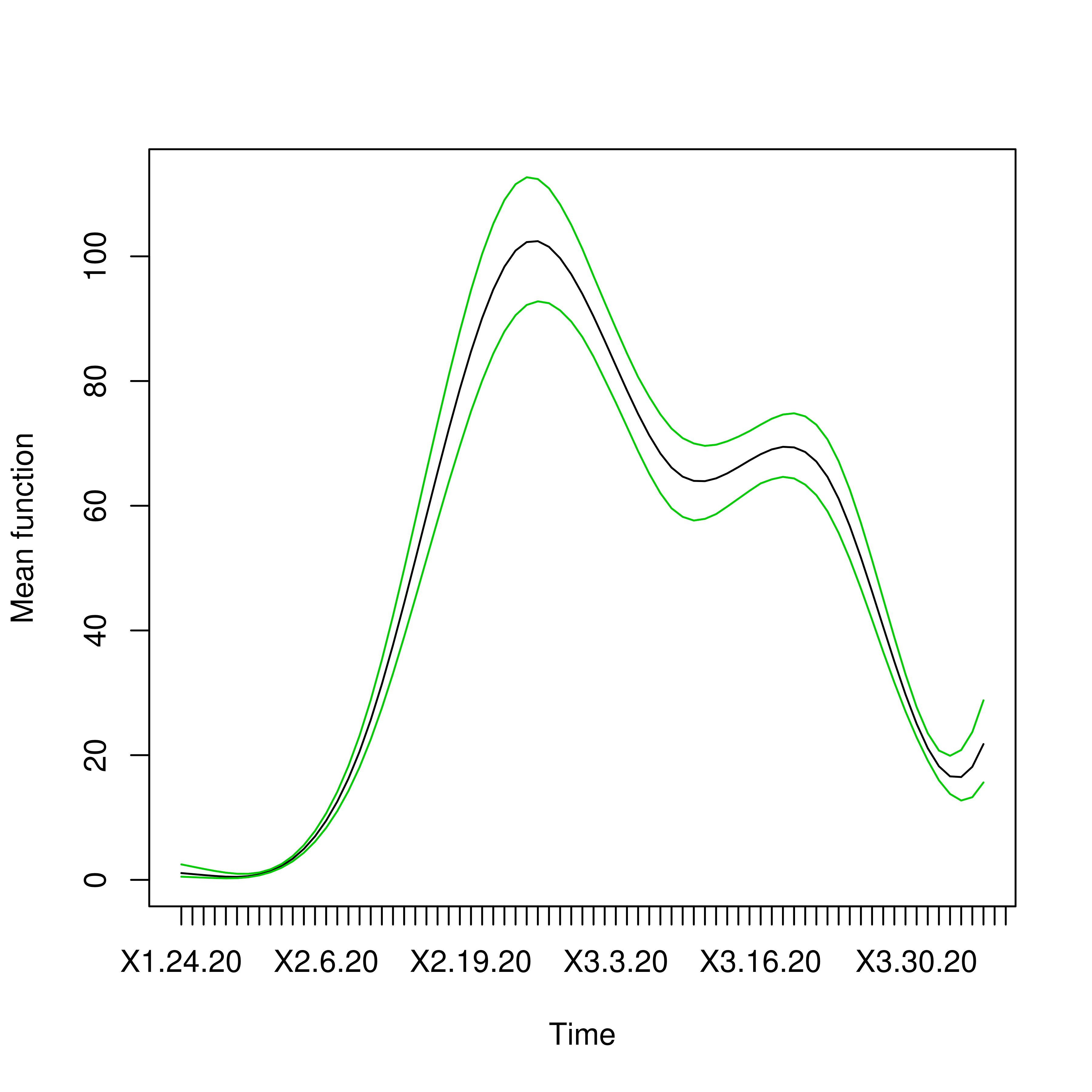}}
		\subfigure[Seoul-$a(\cdot)$ functions]{\label{fig:a.2}\includegraphics[width=60mm]{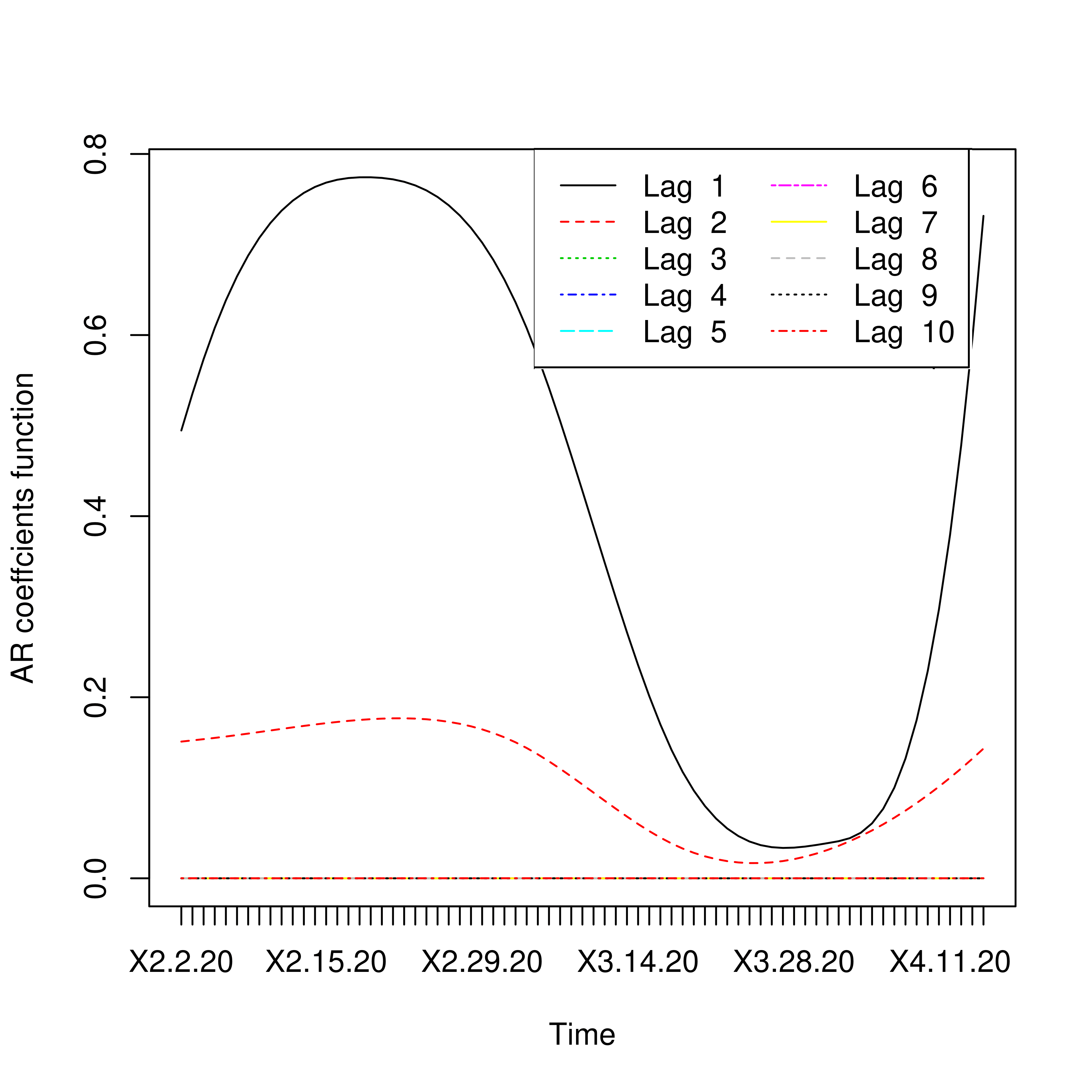}}
		\caption{Estimated mean functions in 1st column and estimated AR coefficient functions in the 2nd column for Seoul, South Korea. Black is the estimated curve along with the 95\% pointwise credible bands in green for the mean function.} 
		\label{fig:SK}
	\end{figure}

	\textit{Intercept trend:} The trend function $\mu(\cdot)$ behaves very similarly for US (Fig \ref{fig:US}), Spain (Fig \ref{fig:Spain}) and Italy (Fig \ref{fig:Italy}. The nature is also prevalent in NYC (Fig \ref{fig:NYC}) however, the same looks somewhat different for Seoul (Fig \ref{fig:SK}) and very different for Hubei (Fig \ref{fig:hubei}). Since this analysis is using the data post Jan 23, whereas the first set of infections in Hubei was in mid-December, it is natural that the peak has occurred sometimes earlier and thus the decreasing trend. The slightly bimodal nature for Seoul looks interesting however this might be due to the relatively smaller numbers for Seoul. One can also find similarity in the number of days to reach a peak after strict lockdown has been enforced. Our estimate puts it around 15-20 days.
	
	\textit{AR(1) function:}
	Note that, at a country level, for each of the US, Spain, and Italy, the estimated lag 1 coefficient $a_1(\cdot)$ were not large and generally did not vary much over time. However, the pattern is very different for the cities as one can see this coefficient dominates the other lags significantly. Hubei and Seoul showed a declining curve with periodic peaks however for NYC, there was an initial decline, possibly due to the low values around February. But since the beginning of March, the $a_1(\cdot)$ grew very steadily which means the infection was spreading in an exponential fashion which matches the empirical numbers we observed in this time-frame. The peak for the mean function $\mu(\cdot)$ is downward at present but the worrying sign is the upward nature of $a_1(\cdot)$. Interestingly one can see since the lockdown was imposed strictly on March 20 in New York, $a_1$ has declined and thus it is not unfair to conclude that this measure has helped in prohibiting the rate of spread. This also explains why we see similar patterns in Hubei and Seoul for the AR(1) function. Also, a natural question is why we do not observe this at a country level. A possible explanation is that the countries are more heterogeneous and with varying degrees of initial infection and local interference.
	
	\textit{Important lag:}
	The TVBARC model for the US and Italy shows interestingly the sixth lag is dominating lag 1 or 2 almost uniformly. We think this is an extremely important find as this matches medical research that talks about the incubation period of the virus. The number of days required for the symptoms to show up after an infection has been spread is currently being widely researched and this incubation day has been proposed to have a median of 6 days and a 98\% quantile of 11 days, see \cite{incubation}. Our finding is coherent with this. This direction of research could potentially be transformative since while estimating the basic reproduction number, there is no result, to the best of our knowledge that could estimate the lag between symptom onset between infector and infectee from the data itself.  
	
\begin{figure}
		\centering
		\includegraphics[width=100mm]{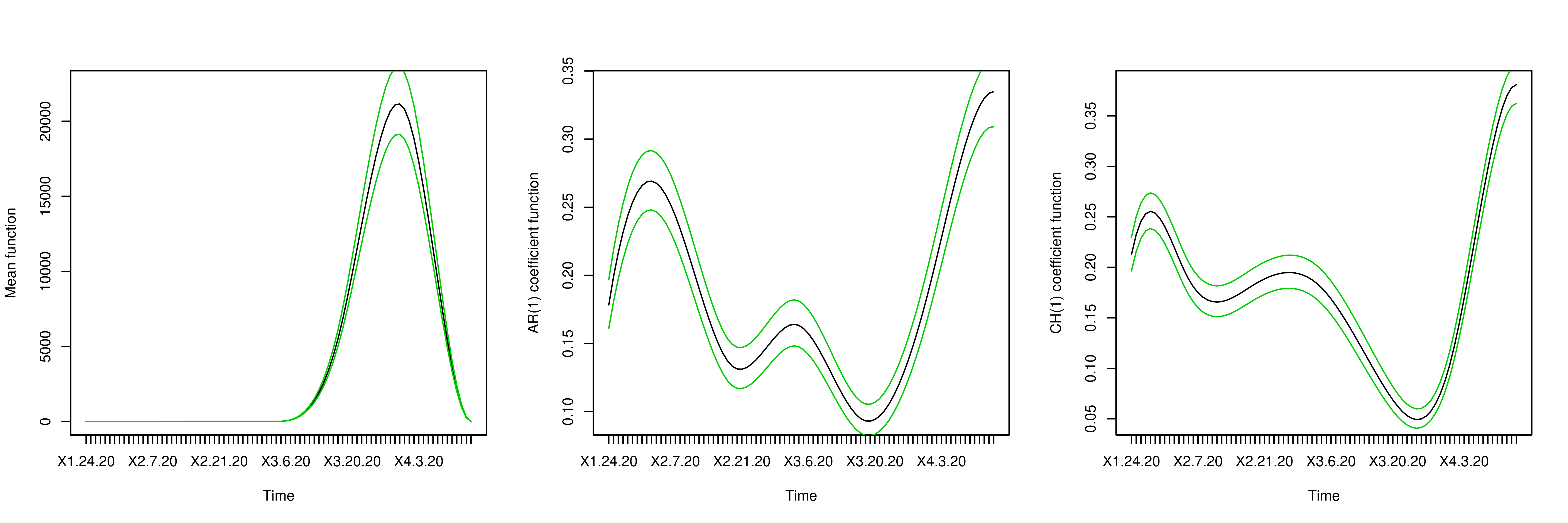}
		\caption{Estimated coefficient functions for the TVBINGARCH(1,1) on USA data. Black is the estimated curve along with the 95\% pointwise credible bands in green.}
		\label{fig:TVBINGARCHUS}
	\end{figure}

\begin{figure}
		\centering
		\includegraphics[width=100mm]{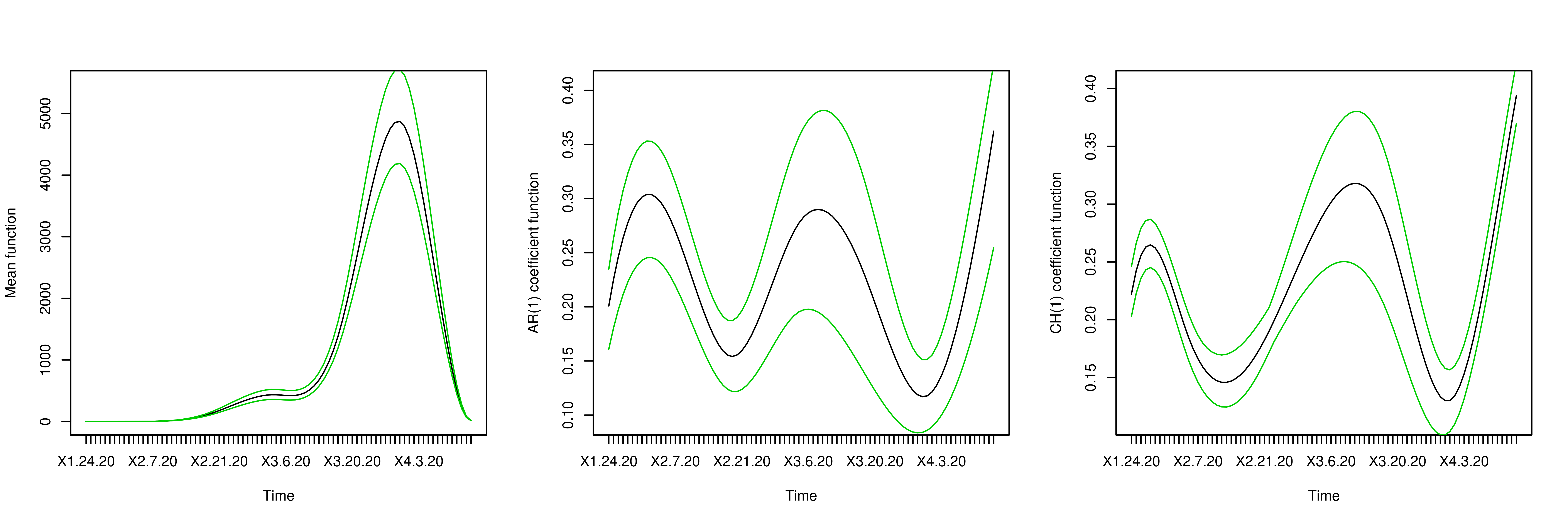}
		\caption{Estimated coefficient functions for the TVBINGARCH(1,1) on Italy data. Black is the estimated curve along with the 95\% pointwise credible bands in green.}
		\label{fig:TVBINGARCHItaly}
	\end{figure}

\begin{figure}
		\centering
		\includegraphics[width=100mm]{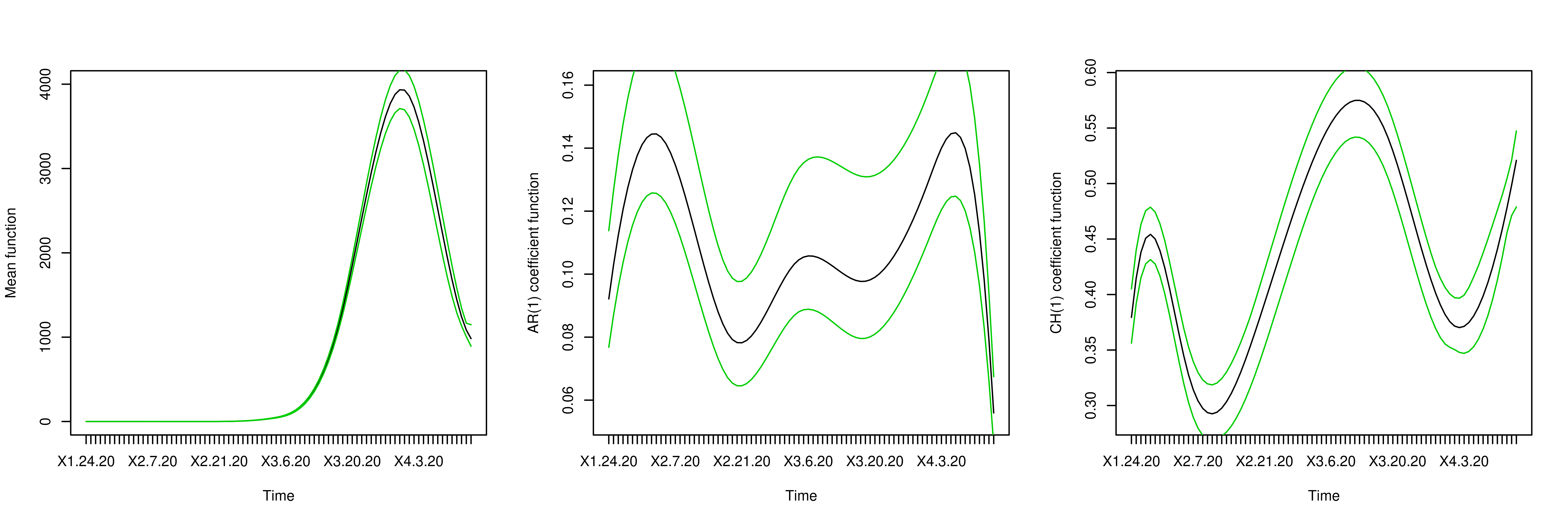}
		\caption{Estimated coefficient functions for the TVBINGARCH(1,1) on Spain data. Black is the estimated curve along with the 95\% pointwise credible bands in green.}
		\label{fig:TVBINGARCHSpain}
	\end{figure}

\begin{figure}
		\centering
		\includegraphics[width=100mm]{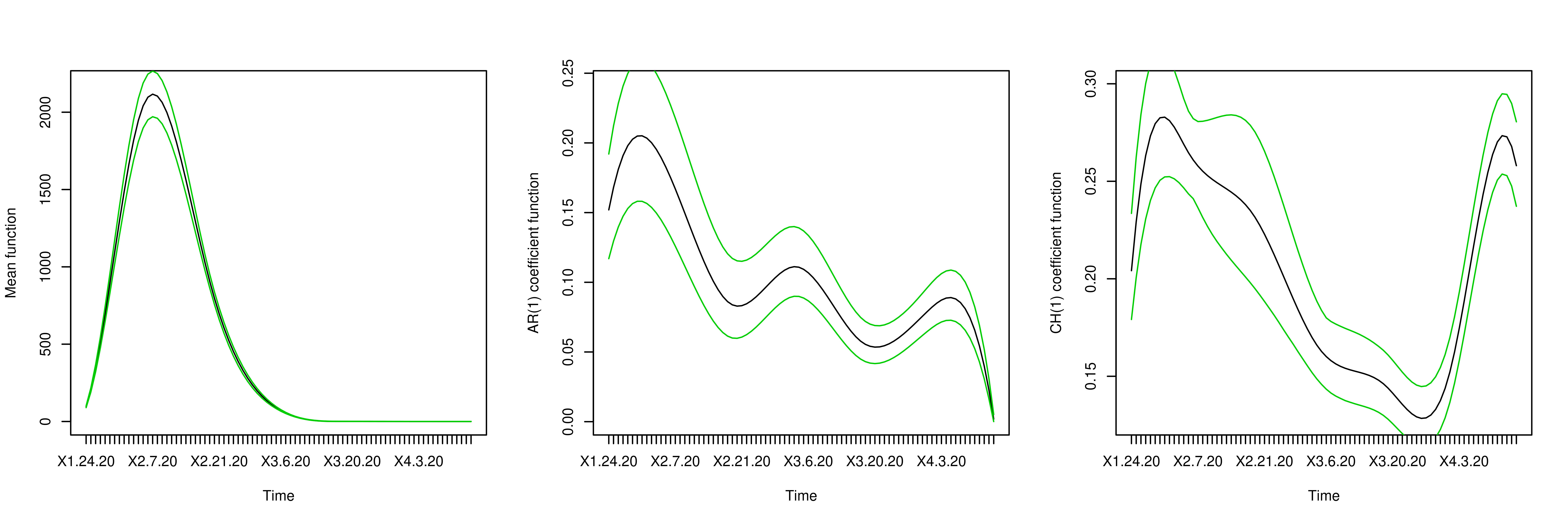}
		\caption{Estimated coefficient functions for the TVBINGARCH(1,1) on Hubei data. Black is the estimated curve along with the 95\% pointwise credible bands in green.}
		\label{fig:TVBINGARChubei}
	\end{figure}

\begin{figure}
		\centering
		\includegraphics[width=100mm]{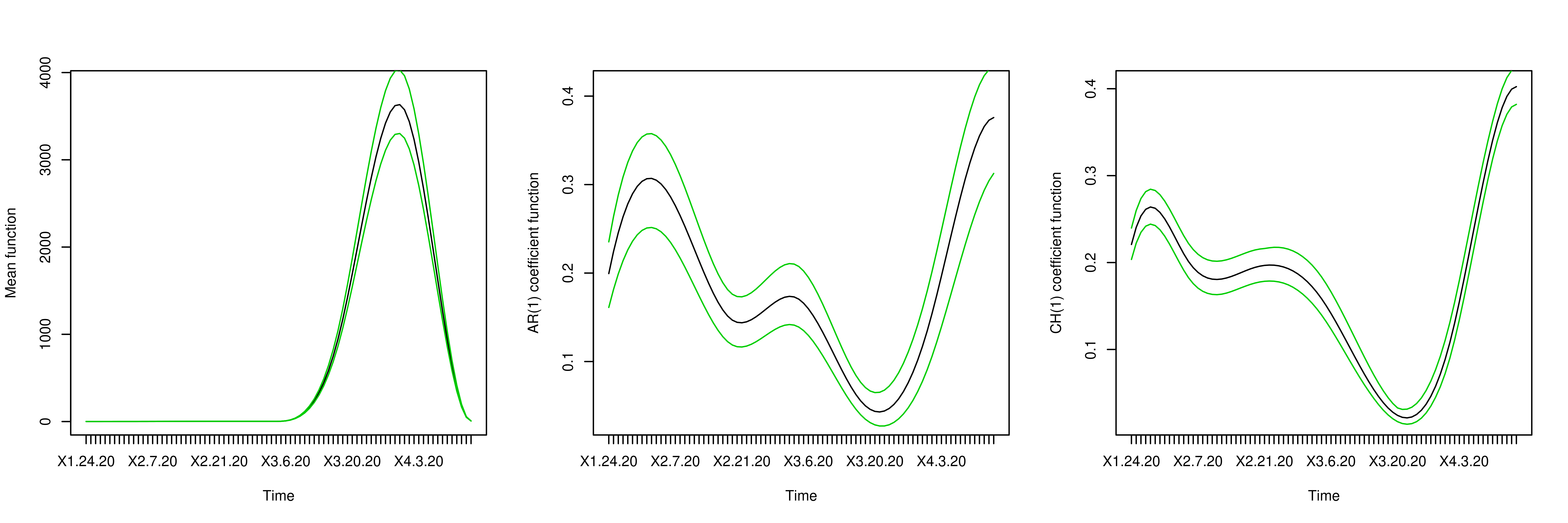}
		\caption{Estimated coefficient functions for the TVBINGARCH(1,1) on NYC data. Black is the estimated curve along with the 95\% pointwise credible bands in green.}
		\label{fig:TVBINGARCHNYC}
	\end{figure}

\begin{figure}
		\centering
		\includegraphics[width=100mm]{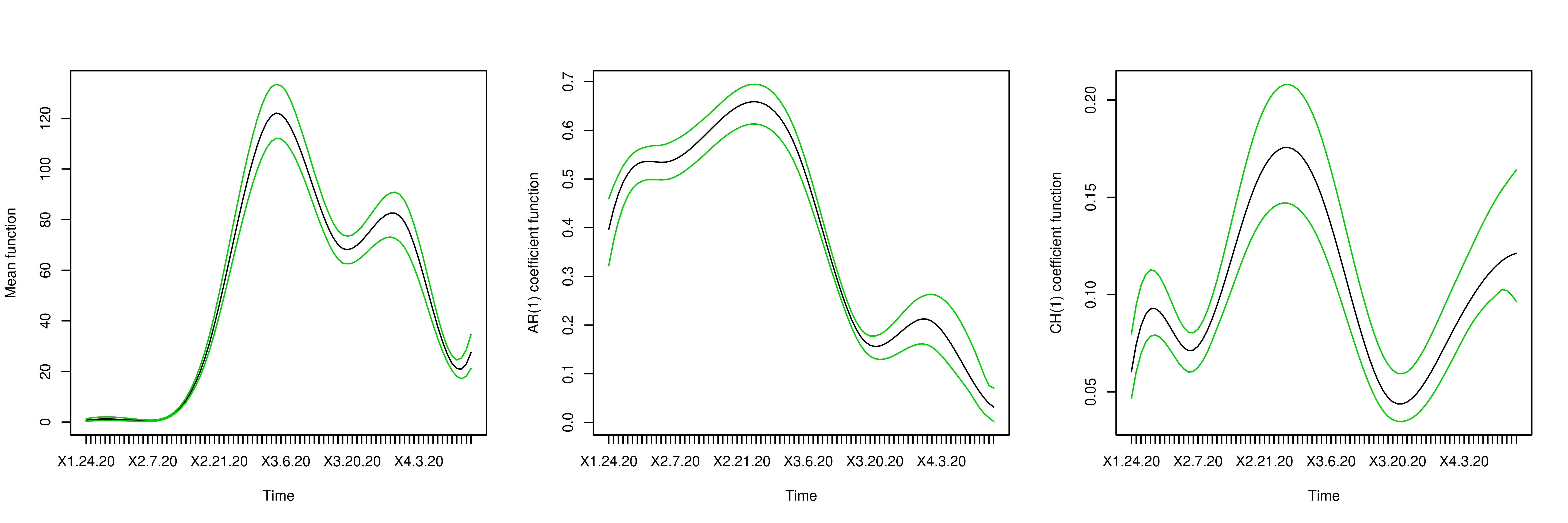}
		\caption{Estimated coefficient functions for the TVBINGARCH(1,1) on Seoul, South Korea data. Black is the estimated curve along with the 95\% pointwise credible bands in green.}
		\label{fig:TVBINGARCHSK}
	\end{figure}

	\textit{TVBINGARCH model:}
	Finally, we conclude our discussion of real data analysis with a very comprehensive model such as INGARCH. Even for INGARCH(1,1) model, the single additional recursive parameter $b_1(\cdot)$ in light of model \eqref{TVBING} allows us for an excellent fit instead of finding the number of lags up to which one should fit. From Fig \ref{fig:TVBINGARCHUS}, \ref{fig:TVBINGARCHItaly} and \ref{fig:TVBINGARCHSpain}, one can see there is a striking similarity between the three countries the US, Spain, and Italy for the AR(1) and CH(1) parameter curves. Hubei (Fig \ref{fig:TVBINGARChubei}) and NYC (Fig \ref{fig:TVBINGARCHNYC}) show similarity while Seoul (Fig \ref{fig:TVBINGARCHSK}) has a different pattern for the CH(1) parameter. Many of these curves have multiple local peaks which corroborate well with the numbers also fluctuating a little bit while flattening out. More interestingly, for all the 6 figures one can see the CH(1) parameter is currently having an upward trend while the intercept trend $\mu(\cdot)$ is going down. This dichotomy is an interesting find and can be corroborated with the fact that for a Poisson random variable the intensity parameter is both the mean and variance. While the mean, in general, is going down the variability for small numbers is somewhat relatively more.
	
\begin{figure}
		\centering
		\includegraphics[width=80mm]{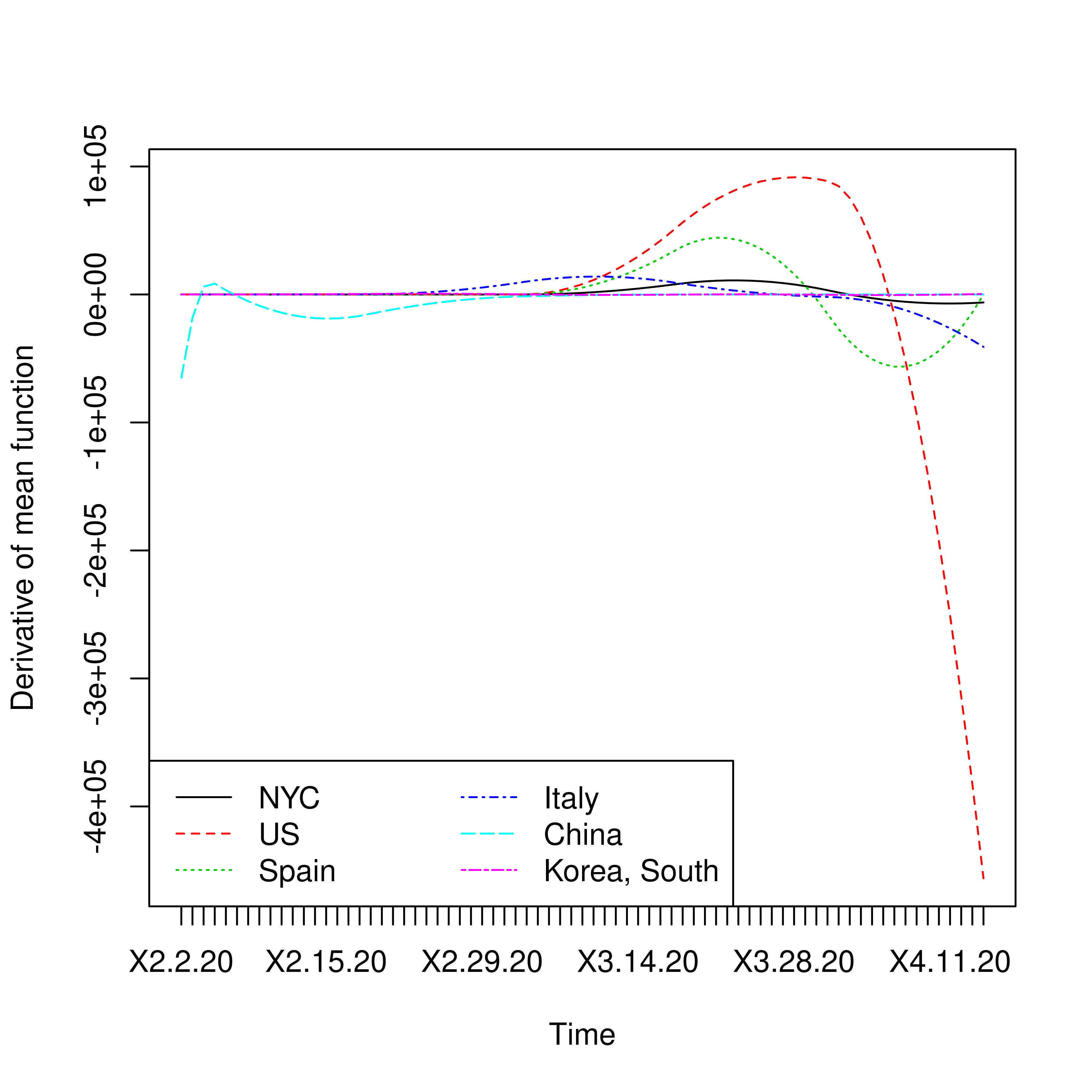}
		\caption{Estimated derivative of the mean coefficient functions for the selected set of regions.}
		\label{fig:muder}
	\end{figure}

	We also provide the estimated derivatives of the estimated $\mu(\cdot)$ functions in Fig \ref{fig:muder}. Overall, we believe that our analysis of data from three countries and three populated cities depicts a comprehensive picture of the mechanism of virus spread. 
	
	\section{Discussion and Conclusion}
	\label{discussion}
	We propose a time-varying Bayesian autoregressive model for counts (TVBARC) and time-varying Bayesian integer-valued generalized autoregressive conditional heteroskedastic model (TVBINGARCH) with linear link function within Poisson to study the time series of daily new confirmed cases of COVID-19. We develop a novel hierarchical Bayesian model that satisfies the stability condition for the respective time-varying models and propose an HMC algorithm based MCMC sampling scheme. The `R' function with an example code can be found at {\url{https://github.com/royarkaprava/TVBARC}}. Relying on the proposed hierarchical Bayesian model, one can develop a time-varying Bayesian model for positive-valued time-series data too. 
	
	We summarize our main findings from the analysis of COVID-19 datasets here. First, we address the time-varying nature of the dataset that takes care of not only how the virus spreads but also different executive restrictions or government interference. It is a difficult task to pour in other covariates as first it is debatable exactly what to include and second different countries, the province even the residents probably behave differently. To keep the flexibility of how the numbers evolve outside the autoregressive effects we choose to keep a mean/intercept coefficient $\mu(\cdot)$. With that model set-up, we analyze three different countries and three cites. We find out interesting similarities between how the $\mu(\cdot)$ behaves over time and see that typically there is a downward trend after around 12 days after lockdown measures have been enforced. However, for the AR(1) coefficient, the trends often show multiple peaks probably due to the asymptomatic spreading capability of the virus. On the same note, another interesting find is to see how we find lag number 6-7th to be important in the majority of the cases conforming to the facts published about the incubation period length of coronavirus. We also propose an INGARCH model which is more comprehensive than just an AR model and observed an interesting phenomenon in both simulation and real-data analysis. We saw that from a predictive perspective we cannot say the INGARCH always dominates an AR model and also when we settle for a small order INGARCH model we tend to lose out on the interesting 6-7th lag phenomenon that is prevalent with this disease for many cities and countries. However, time-varying INGARCH can be useful in summarizing the coefficients more compactly. 
	
	There is growing skepticism in the various finding of R0, the basic reproduction number of the pandemic. These are derived from the popular SIR model but, due to the huge non-stationary propagation of the data, it is heavily dependent on start and end time. Also, the lags for symptom onset between an infector and infectee is difficult to estimate and often done not from the data itself, but  SARS and MERS. We do not think this is a correct approach since the dynamics of virus spread for COVID-19 has been different. Thus we offer a different approach rather than having R0 in our model. However if one wants to make an explicit connection, one can propose the `time-varying R0' model extending from \cite{R0packagepaper} as, $$X_t|\sF_{t-1} \sim \mathrm{Poi}(\lambda_t), \lambda_t = R(t)\sum_{i=1}^{\infty} X_{t-i}w_i $$ one can see this is very similar to saying $a_i(t/T)=R(t)w_i$. Moreover, the interesting find in our paper allows us to highlight that the weights $w_i$ have a high concentration around lag 6. This alternative way of using the data itself to re-estimate the reproduction number can be transformative.

	As future work, it will be interesting to include some country-specific information such as demographic information, geographical area, the effect of environmental time-series, etc in the model. These are usually important factors for the spread of any infectious disease. We can also categorize the different types of government intervention effects to elaborate more on the specific impacts of the same. In the future we wish to analyze the number of deaths, number of recovered cases, number of severe/critical cases, etc. for these diseases as those will hopefully have different dynamics than the one considered here and can provide useful insights about the spread and measures required. For computational ease, we have considered same level of smoothness for all the coefficient functions. Fitting this model with different levels of smoothness might be able to provide more insights. Other than building time-varying autoregressive models for positive-valued data using the hierarchical structure from this article, one interesting future direction is to extend this model for vector-valued count data. In general, it is difficult to model multivariate count data. There are only a limited number of methods to deal with multivariate count data \citep{besag1974spatial,yang2013poisson,roy2019nonparametric}. Building on these multivariate count data models, one can extend our time-varying univariate AR$(p)$ to a time-varying vector-valued AR$(p)$. On the same note, even though we imposed Poisson assumption for increased model interpretation, in the light of the upper bounds for the KL distance, it is not a necessary criterion and can be applied to a general multiple non-stationary count time-series. Extending some of the continuous time-series invariance results from \cite{karmakar2020optimal} to multiple count time-series will be an interesting challenge. Finally, we wish to undertake an autoregressive estimation of the basic reproduction number with the time-varying version of compartmental models in epidemiology immediately. 
	
	\bibliographystyle{bibstyle}
	\bibliography{main}

\end{document}